\documentclass[a4paper,10pt]{article}
\usepackage[hmargin=1.5cm,vmargin=1.5cm]{geometry}
\usepackage{booktabs}
\usepackage{dsfont}
\usepackage{amsmath, amssymb}

\usepackage{amsfonts}
\usepackage{float}
\usepackage[small,it]{caption}
\usepackage{multirow}
\usepackage{setspace}
\usepackage{pdfpages}
\usepackage{exscale,relsize}
\usepackage{subfig}
\usepackage{setspace}
\usepackage{empheq}
\usepackage[numbers,sort]{natbib}
\usepackage{comment}
\usepackage{tcolorbox}
\usepackage{graphicx}
\usepackage{epstopdf}
\usepackage{soul,xcolor}
\usepackage{lineno}
\soulregister\cite7
\soulregister\ref7
\soulregister\pageref7

\begin{document}
\title{Evolution of longevity, age at last birth and sexual conflict with grandmothering}
\author{Matthew H.\ Chan$^{1}$,
\and Kristen Hawkes$^{2}$,
\and Peter S.\ Kim$^{3}$}
\date{\today}
\maketitle

\footnotetext[1]{School of Mathematics and Statistics, University of Sydney,
 NSW 2006, Australia. {\tt M.Chan@maths.usyd.edu.au}}
 \footnotetext[2]{Department of Anthropology, University of Utah, Salt Lake City,
 UT 84112, USA. {\tt Hawkes@anthro.utah.edu}}
   \footnotetext[3]{School of Mathematics and Statistics, University of Sydney,
 NSW 2006, Australia. {\tt PKim@maths.usyd.edu.au}}
\renewcommand{\thefootnote}{\arabic{footnote}}
\newenvironment{acknowledgements}{{\flushleft \bf Acknowledgements:}}{}

\begin{abstract}
We use a two-sex partial differential equation (PDE) model based on the Grandmother Hypothesis. We build on an earlier model by Kim et al. \cite{116} by allowing for evolution in both longevity and age at last birth, and also assuming that post-fertile females support only their daughters' fertility. Similarly to Kim et al. \cite{116}, we find that only two locally stable equilibria exist: one corresponding to great ape-like longevities and the other corresponding to hunter-gatherer longevities. Our results show that grandmothering enables the transition between these two equilibria, without extending the end of fertility. Moreover, sensitivity analyses of the model show that male competition, arising from a skew in the mating sex ratio towards males, plays a significant role in determining whether the transition from great ape-like longevities to higher longevities is possible and the equilibrium value of the average adult lifespan. Whereas grandmothering effects have a significant impact on the equilibrium value of the average age at last birth and enable the transition to higher longevities, they have an insignificant impact on the equilibrium value of the average adult lifespan.
\end{abstract}

{\bf Key-words}: Population dynamics, Evolutionary dynamics, Grandmother Hypothesis, Postmenopausal longevity.

\pagebreak
\section{Introduction}

A key feature that distinguishes humans from our closest evolutionary cousins is a much higher longevity; for example, the marked increase in age-specific mortality in the common chimpanzee starts at approximately 20 years, whereas in hunter-gatherer populations the rise occurs at about 45 years and is noticeably less steep \cite{120}. However, for both humans and other great apes, the average end of female fertility is around 45 years \cite{123,124}. A consequence of this is that humans are the only primate to have a significant proportion of adult females living beyond their fertile years \cite{126,127,154}. Although it is widely assumed that survival past menopause is only a recent phenomenon due to improvements in life expectancy from medical and public health technology, this assumption is based on a misunderstanding. Life expectancy is an average that is strongly affected by high infant and juvenile mortality rates. Data on historical and hunter-gatherer populations show that females who survived past their juvenile years are much more likely than not to survive past menopause, even though life expectancies in these populations are less than 40 years \cite{130}. Moreover, in classic model life tables built from national census data, a tripling of life expectancy (from 20 to 60 years) has a trivial effect on the proportion of elders in the population \cite{133,153}. As noted by Levitis et al. \cite{126}, ``even the worst surviving human population has Postreproductive Representation higher than the highest recorded value for nonhuman primates in protected environments."
\\
\\
The human mismatch between the end of fertility and longevity is seemingly paradoxical as natural selection generally favours traits that increase life-time reproduction. However, there is evidence from hunter-gatherers of grandmothers supplying foods that just weaned juveniles cannot acquire effectively for themselves \cite{155,156}. In other mammals, including other great apes, weaned juveniles can feed themselves, while humans continue to depend on provisioning after weaning. The help of older women allows their daughters to have next offspring while the previous one is still dependent. This prompted the ``Grandmother hypothesis" to explain the evolutionary origins of our distinctive post-menopausal life stage \cite{157,158}. The hypothesis proposes increased longevity was favoured in our lineage without an increase in the age female fertility ends when ancestral populations began to rely on foods weaned juveniles could not handle for themselves. Under those ecological circumstances, mothers would have to feed their offspring longer, but with subsidies from grandmothers they could have next offspring sooner; and longer-lived grandmothers, without infants themselves, could support more grandchildren. Moreover, the increase in both body size and delay in maturation age are expected consequences of greater longevity according to Charnov's model of mammalian life history, which aims to explain the ways female life histories vary across the mammals \cite{129}. In Charnov's mammal model, age at independence, age at maturity and adult body size all increase in predictable ways with increases in adult lifespan.
\\
\\
In this study, we build upon a mathematical model by Kim et al. \cite{117} and a more sophisticated agent-based model by Kim et al. \cite{116}. The Kim et al. \cite{116} model tracks the evolutionary trajectory in longevity, while fixing the age at last birth based on the similarity in oldest ages at parturition in humans and other great apes. While that model addresses the question ``could grandmothering drive the evolution of increased longevity while keeping the end of female fertility fixed?", it leaves unanswered the question why female fertility ends at about 45 years. We formulate a partial differential equation (PDE) model which shares many of the same assumptions as Kim et al. \cite{116}, but allows for evolution in age at last birth, restricts grandmothers to support only their daughters and uses a more realistic mortality function which in turn generates a more realistic age structure at both the great ape-like equilibrium and hunter-gatherer equilibrium.
\\
\\
{The aim of the study is to determine whether there can even exist an evolutionary trajectory from a great ape-like equilibrium to a hunter-gatherer-like equilibrium under reasonable parameter settings, many of which are based on empirical data (the exceptions being the male and female fertility-longevity tradeoff functions, see Sections \ref{femaletradeoffsect} and \ref{maletradeoffsect} for details). We note that in our study we require both the great ape-like and hunter-gatherer-like equilibria to match empirical data on life expectancy, age at first birth, age at last birth and lastly age profile of wild chimpanzees and hunter-gathers respectively. As such, our definitions for both of these equilibria are more stringent than in previous studies which examine the evolutionary effects of grandmothering such as by Kim et al. \cite{116}, Kim et al. \cite{117}, Pavard \& Branger \cite{118} and Kachel \cite{115}. We show that such an evolutionary trajectory exists, whereby grandmothering is the sole mechanism responsible for the transition between the two equilibria. Moreover, in Section \ref{sensitivitysect} we show that the existence of such an evolutionary trajectory is not automatic, but is instead sensitive to the parameters of the model.}

\section{Model}

We consider an age-structured system with two compartments, male and female population densities, denoted by $u_m(a,M,L,t)$ and $u_f(a,M,L,t)$ respectively, where $a \in [0,\tau_3]$ denotes age, $M \in [0,1]$ is a measure of the age at last birth, $L \in [0,1]$ is a measure of an individual's longevity and $t$ denotes time. {The population age-structure and mortality dynamics are given by the McKendrick-von Foerster equations}

\begin{equation}
\label{mckendrickpdegm}
\begin{cases}
\frac{\partial u_f}{\partial t} = -\frac{\partial u_f}{\partial a} -\mu(a,L) u_f, \\
\frac{\partial u_m}{\partial t} = -\frac{\partial u_m}{\partial a} -\mu(a,L) u_m,
\end{cases}
\end{equation}
\\
{where $\mu(a,L)$ is the age-specific mortality rate for an individual with longevity $L$.
\\
\\
The boundary condition for system (\ref{mckendrickpdegm}) governs several mechanisms of the model: whether individuals are eligible to mate, how individuals mate, whether a female has a mother who is alive and eligible to grandmother, the benefit from grandmothering effects and how offspring inherit their parent's longevity and age of last birth trait values. We first explain the assumptions behind these mechanisms and then introduce the mathematical formulation of the boundary condition in Section \ref{bcsect}.}

\subsection{Fertility}
\label{fertsect}
We assume all individuals simply progress from birth through to sexual maturity. Unlike Kim et al. \cite{116}, we do not include stages of nursing, weaned dependency or independence for simplicity. Females are fertile between ages $a=\tau_1(L)$ and $\tau_2(M)$ respectively, where

\begin{equation}
\label{tau1}
\tau_1(L) = \frac{1}{3} \left( 35L+29 \right)
\end{equation}
\\
and

\begin{equation}
\label{tau2}
\tau_2(M) = 20+55M.
\end{equation}
\\
Thus, the age of female sexual maturity ranges from $9.6$ to $21.3$ and the age at last birth ranges from $20$ to $75$, depending on trait values $M$ and $L$. We design these functions such that $(M,L) = (0.455,0.2)$ and $(M,L) = (0.455,0.8)$ correspond to great ape- and hunter-gatherer-like values respectively. This translates to great apes having a female sexual maturity of 12 years and humans having a female sexual maturity of 19 years, with both having an age at last birth of 45 years. These approximately match the data in Sugiyama \cite{148}, Hill \& Hurtado \cite{142}, Walker et al. \cite{149} and Hawkes et al. \cite{124}. We intentionally let $\tau_1(L)$ and $\tau_2(M)$ have unrealistically large ranges to investigate the $M$ and $L$ trait values selected by the population. The form for both $\tau_1(L)$ and $\tau_2(M)$ were chosen such that they are linear within the ranges. We let the age of frailty for both males and females to be $\tau_3 = 75$. As with Kim et al. \cite{116}, we assume males compete for paternities from age $15$ until frailty, regardless of their $M$ or $L$ traits. Kim et al. \cite{116} base this on data from living great apes \cite{145,146}.

\subsection{Mortality}
\label{mortsect}

\begin{figure}[H]
  \hspace{1mm}\centerline{
  \subfloat[]{\label{muplot}\includegraphics[scale=0.5]{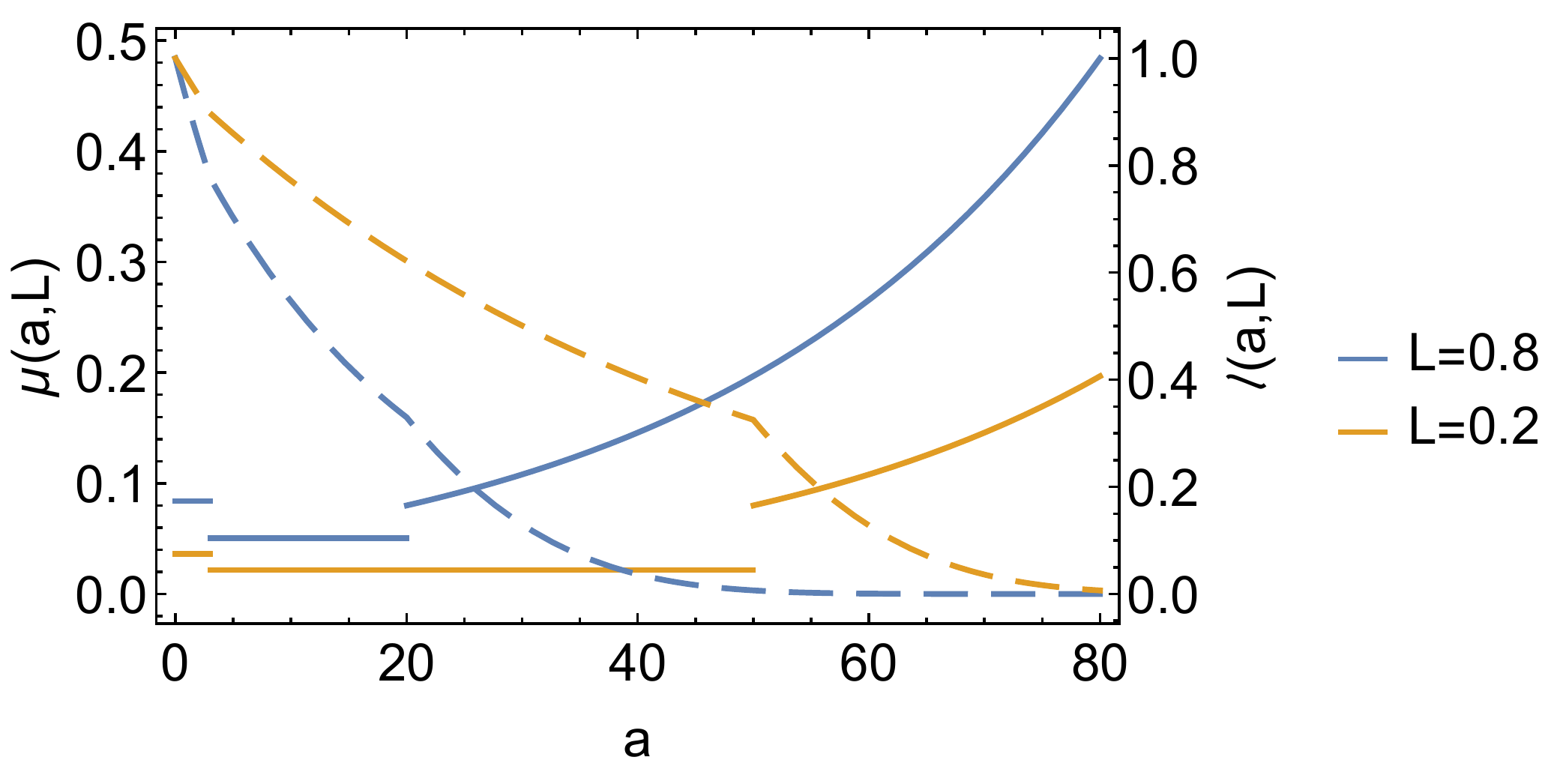}}
  \hspace{1mm}
  \subfloat[]{\label{e0plot}\includegraphics[scale=0.5]{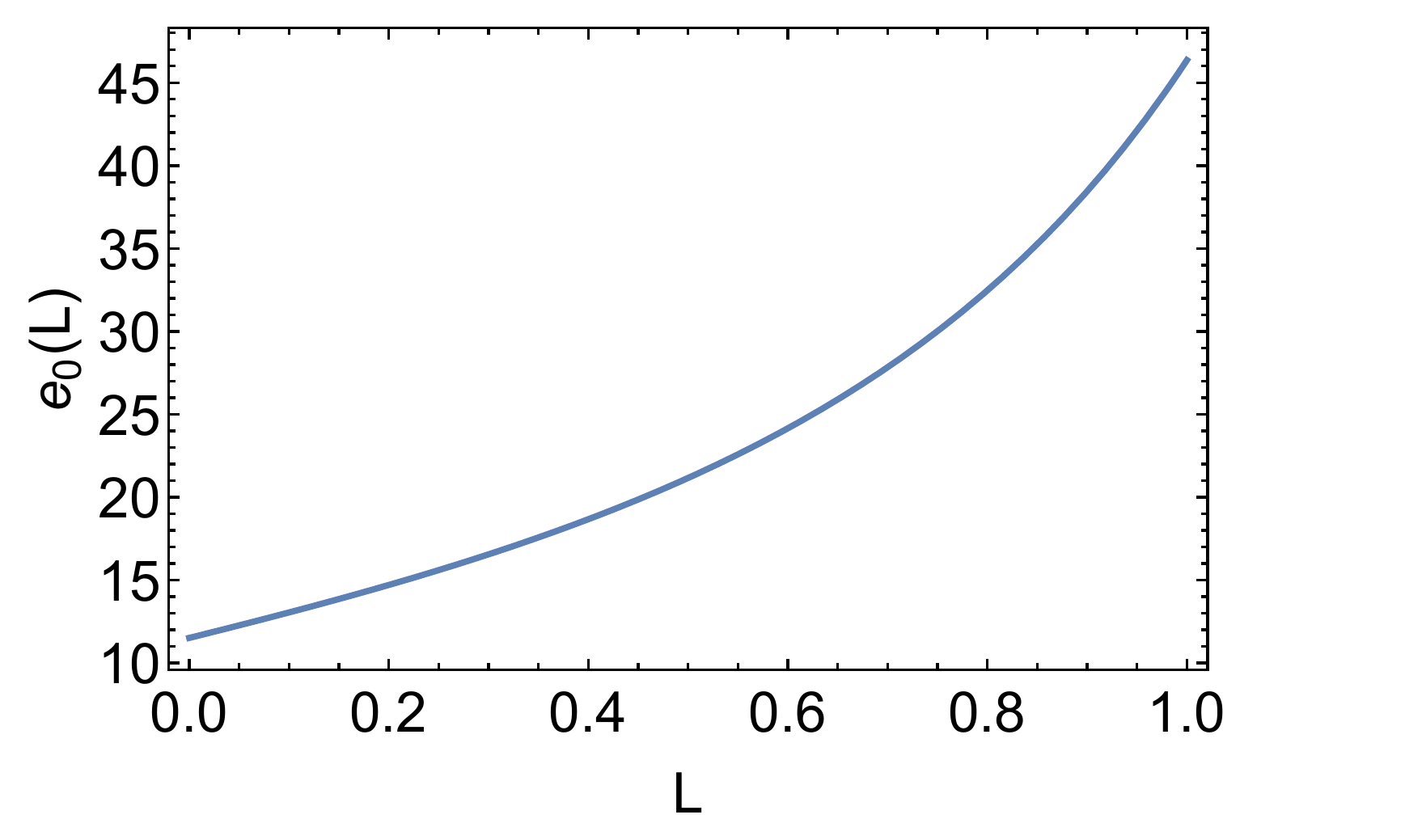}}}
  \caption{Solid and dashed lines in plot (a) show the mortality function $\mu(a,L)$ and survivorship function $l(a,L)$ respectively for $L=0.2$ and $L=0.8$. Plot (b) shows the life expectancy of a newborn $e_0(L)$ varying with $L$.}
  \label{plots0}
\end{figure}

All individuals have a mortality rate $\mu(a,L)$ dependent on age $a$ and longevity $L$, where there are three distinct stages: early, middle and late mortality stages. This serves as a simplified model of the age-specific mortality data shown below in Figure \ref{hawkesmortality}. Distinguishing early, middle and late stages allows for a more realistic mortality function than the constant mortality rate assumed by Kim et al. \cite{116}. The form and parameters of the mortality function are chosen such that the survivorship curves and age-specific mortality rates generated in Figure \ref{muplot} are consistent with survivorship data in Gurven \& Kaplan \cite{130} and age-specific mortality data in Figure \ref{hawkesmortality}. We assume that the early mortality stage lasts three years regardless of $L$, but the length of the middle stage increases with increasing longevity. The mortality rates in both of these stages are constant for simplicity. We define $\mu(a,L)$ to be


\begin{equation}
\label{mortalityfunc}
\mu(a,L) = 
\begin{cases}
\mu_j (1- 0.8 L) \quad&\text{ if} \quad a \leq 3, \\
\mu_a (1- 0.8 L) \quad &\text{ if} \quad 3 < a \leq a_2(L), \\
\mu_s e^{0.03 (a-a_2(L))} \quad &\text{ if} \quad a > a_2(L),
\end{cases}
\end{equation}
\\
where $a_2(L) = 50L+10$ and $\mu_j$, $\mu_a$ and $\mu_s$ are constants given in Table \ref{lifehistoryparatable}. However, we choose the rate of increase in late mortality to underestimate the age-specific mortality data shown in Figure \ref{hawkesmortality}, as this generates fatter tails which matches survivorship data in Gurven \& Kaplan \cite{130} and generates PRr values closer to those found by Levitis \& Lackey \cite{140}. We note that since $L=0.2$ and $L=0.8$ are designated to correspond to great ape- and hunter-gatherer-like longevity values respectively (see Section \ref{fertsect}), the parameters in Eq.~(\ref{mortalityfunc}) were also chosen such that $e_0(0.2) = 14.7$ and $e_0(0.8) = 32.5$ respectively, which approximately matches hunter-gatherer and wild chimpanzee life expectancy estimates in Gurven \& Kaplan \cite{130} and Hill et al. \cite{144}.
\\
\\
From Eq.~(\ref{mortalityfunc}), we are able to compute the life expectancy $e_0$ of a newborn with trait $L$, by

\begin{equation}
\label{lifeexpect}
e_0(L) = \int_0^\infty \exp \left( -\int_0^a \mu(x,L) \, dx \right) \, da,
\end{equation}
\\
and also the Post-reproductive Representation (PRr) of the population with mean trait values $M$ and $L$, by

\begin{equation}
\text{PRr}(M,L) = \frac{\gamma(\tau_M)}{\gamma(\tau_B)},
\end{equation}
\\
where $\gamma(x)$ is the expected number of years greater than age $x$ a newborn lives, given by

\begin{equation}
\gamma(x) = \int_x^\infty \exp \left( -\int_0^a \mu(y) \ dy \right) \ da
\end{equation}
\\
and $\tau_M$ and $\tau_B$ are the ages at which 95\% of births and 5\% births have been realised respectively (for details on PRr see Levitis \& Lackey \cite{140}). Due to the birth function being independent of age, we have that $\tau_M = 0.95\tau_2(M)$ and $\tau_B=1.05\tau_1(L)$. Note that $\gamma(x)$ is different from the life expectancy given survival to age $x$ (denoted by $e_x$ in the literature) as there is no assumption of survival to age $x$. Fixing $\tau_1(L) = 19$ and varying $\tau_2(M)$ within the interval $[40,50]$ corresponds to the PRr ranging from $0.449$ to $0.235$. For $\tau_M$ close to 40 years, the PRr values obtained matches those of Levitis \& Lackey \cite{140}, who report a PRr value ranging from 0.42 to 0.48 for hunter-gatherers.

\begin{figure}[H]
	\centerline{\includegraphics[scale=0.5]{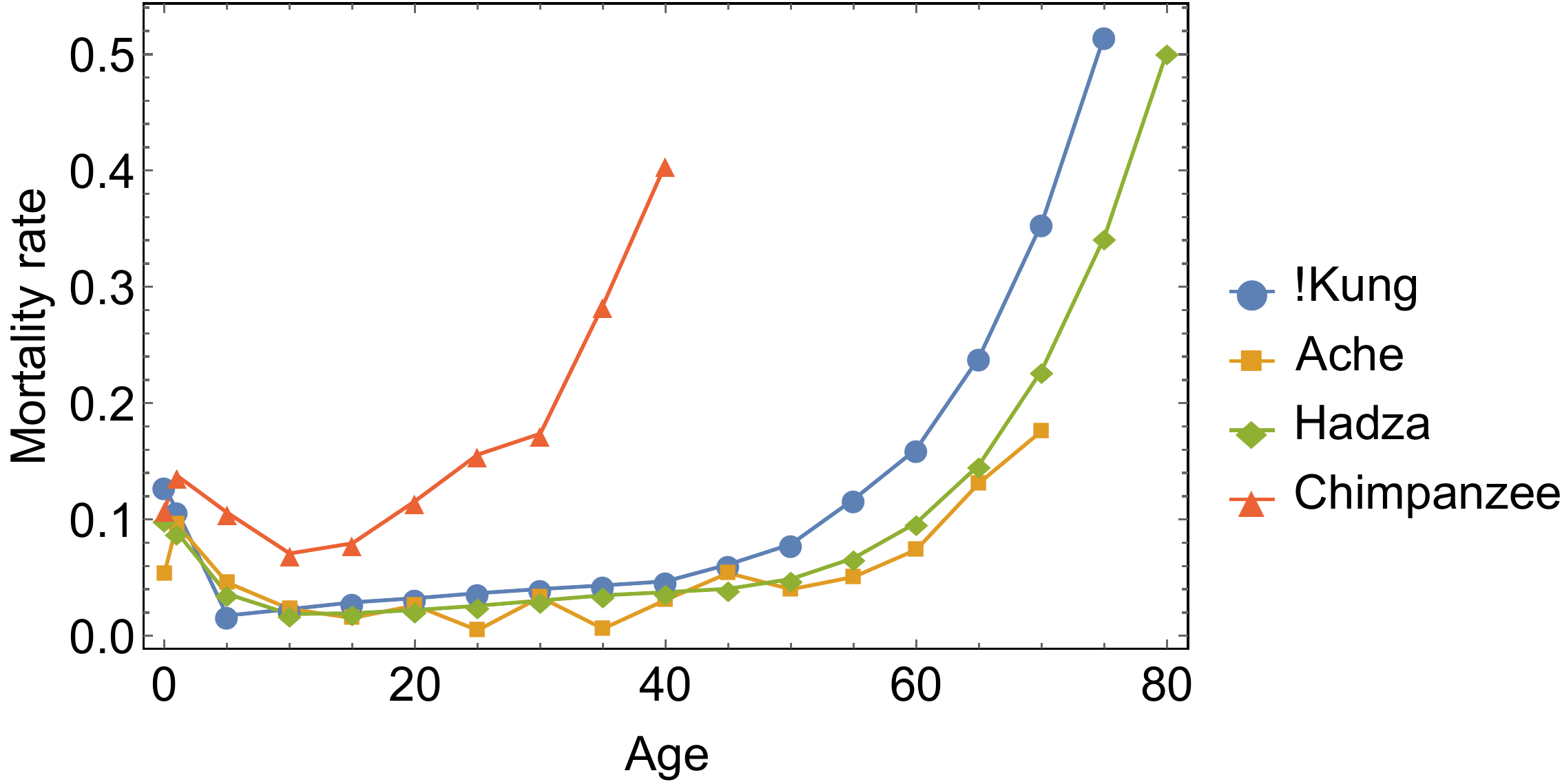}}
  \caption{Age-specific mortality rates for three hunter-gatherer populations (!Kung, Ache, Hadza) and wild chimpanzees. !Kung data from Howell \cite{141}, Ache data from Hill \& Hurtado \cite{142}, Hadza data from Blurton Jones et al. \cite{143} and chimpanzee data (smoothed) from Hill et al. \cite{144}.}
  \label{hawkesmortality}
\end{figure}

\subsection{Mating and offspring}
\label{matingoffsect}

We assume that females randomly select a male to mate with, whereby the resultant offspring inherits the mean of the parents' $M$ and $L$ values, with the possibility of a mutational shift. To express this in a continuous time and population density framework, females mate with the normalised distribution of sexually mature males and give birth to a distribution of offspring in $M$ and $L$. For computational purposes, we assume that $5\%$ of all conceptions will have mutations, where there is a uniform probability of shifting $\pm 0.05$ in either $M$ or $L$.
\\
\\
To keep the population bounded, the number of offspring born is regulated by the total number of females in the system at time $t$, denoted by $\Phi(t)$, via multiplying $b(M,L)$ by $\frac{1}{1+\phi(t)}$. This is given by

\begin{equation}
\Phi(t) = \int_0^{\tau_3} \int_0^1 \int_0^1 u_f(a,M,L) dL dM da.
\end{equation}
\\
{A central idea in life-history theory is that of costs of reproduction, where limited energy reserves imply a tradeoff between current reproductive effort and survivorship or other major life-history traits. We assume that such a tradeoff holds for both male and female individuals, and we give these tradeoff functions specific forms as to ensure their shapes allow for an evolutionary trajectory from a great ape-like equilibrium to a hunter-gatherer equilibrium. We discuss the importance of these tradeoff functions in the Discussion section and explore different forms of the male fertility-tradeoff function in Section \ref{sensitivity_male}.}

\subsubsection{{Female fertility-longevity tradeoff}}
\label{femaletradeoffsect}
In the absence of grandmothers, females with traits $M$ and $L$ have a constant birth rate of $b(M,L)$/year for any age in which the female is fertile, where $b(M,L)$, the averaged reproductive rate, is given by 

\begin{equation}
\label{birthrate}
b(M,L) = \frac{5}{e_0(L)}e^{-0.12 M}e^{-0.33 L}.
\end{equation}
\\
Using $(M,L) = (0.455,0.2)$ and $(M,L) = (0.455,0.8)$ as the great ape-like and hunter-gatherer $M$ and $L$ values respectively (see the Sections \ref{mortsect} and \ref{fertsect} for justification), individuals with great ape-like longevities have a birth rate of $b(0.455,0.2) = 0.3$/year and individuals with hunter-gatherer longevities (without the aid of grandmothers) have a birth rate of $b(0.455,0.8) = 0.11$/year. The great ape-like birth rate matches the data in Sugiyama \cite{148} for chimpanzees aged 16-23 years, where we are including infants which die within three years in the calculation. The hunter-gatherer birth rate of 0.11/year (without a grandmother) is approximated by assuming that fertile females cannot conceive if they have a dependent child. Assuming that children become independent at the age of 8 years, and that conception plus gestation takes 1 year, the reciprocal of this period between births yields 0.11/year.
\\
\\
{We note that we have designed Eq.~(\ref{birthrate}) such that not only does it match empirical data on wild chimpanzee and hunter-gatherer birth rates at $(M,L) = (0.455,0.2)$ and $(M,L) = (0.455,0.8)$ respectively, but also such that a population without grandmothering with $(M,L) = (0.455,0.8)$ has a net reproductive rate less than 1. We find that these conditions require an exponentially decreasing birth rate with respect to $L$. Moreover, we have based the form of Eq.~(\ref{birthrate}) from the interbirth interval of Kim et al. \cite{116} roughly corresponding to $\frac{5}{e_0(L)}$; however, due to different mortality functions used (Kim et al. \cite{116} uses a constant mortality rate) this form for $b(L)$ is not feasible for our model. We include the exponential terms to guarantee that a population with mean trait values corresponding to hunter-gatherers $(M,L) = (0.455,0.8)$ cannot survive without grandmothering.}

\subsubsection{{Male fertility-longevity tradeoff}}
\label{maletradeoffsect}
We model a cost of increased longevity for males by assigning a weighting function $\phi(L)$ that represents the relative probability that a male will outcompete others for a chance at paternity. Similarly to Kim et al. \cite{116}, we base this on Williams' \cite{119} deductions about the effects of natural selection on senescence, that ``successful selection for increased longevity should result in decreased vigour in youth". This is given by

\begin{equation}
\label{malecomp}
\phi(L) = \left(2.3 e^{-\rho L}+1\right)\exp\left( - 0.65 L^{5} \right)/2.3.
\end{equation}

\begin{figure}[H]
  \centerline{\includegraphics[scale=0.5]{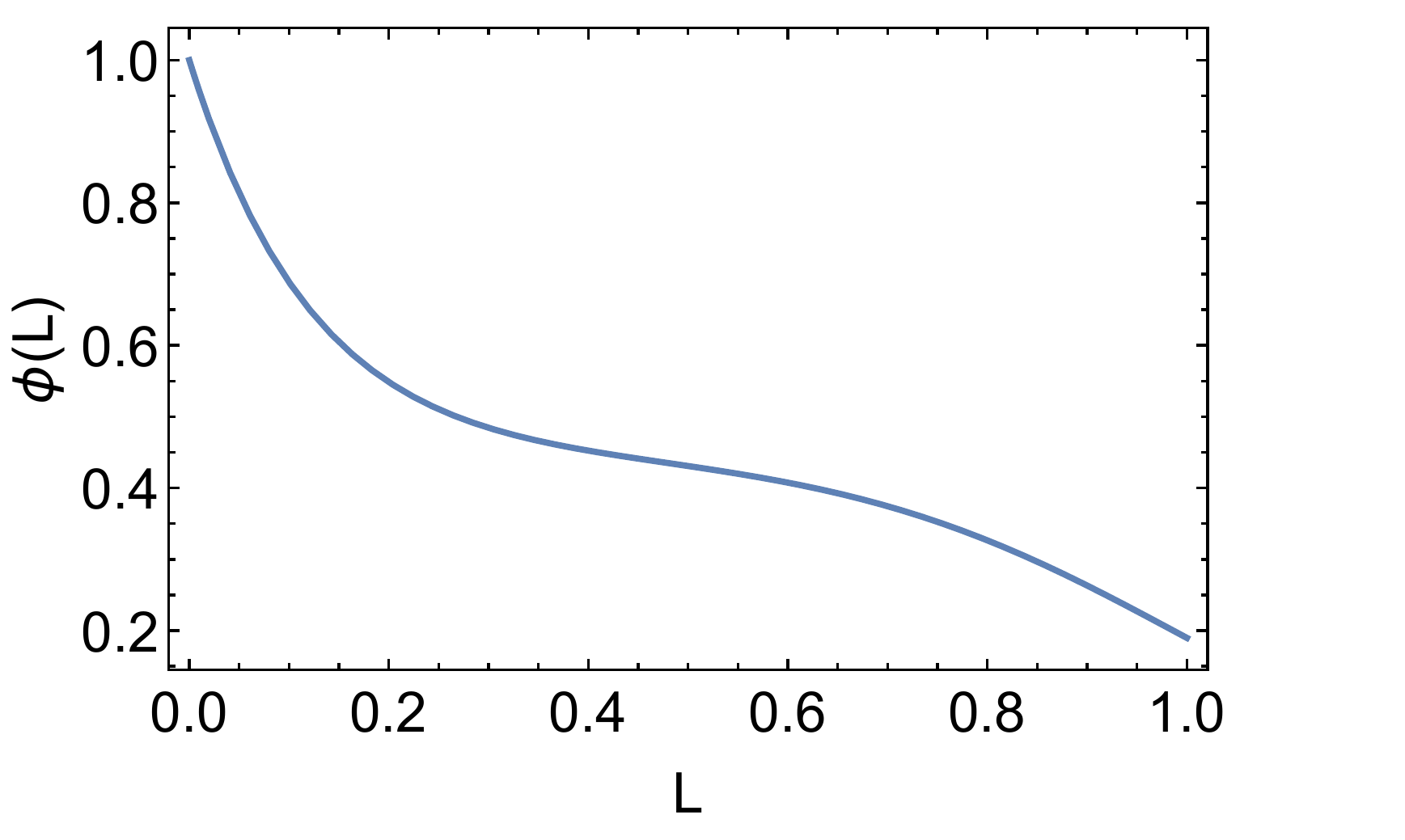}}
  \caption{Male fertility-longevity tradeoff function $\phi(L)$.}
  \label{malecompplot}
\end{figure}

\noindent {Unlike the female fertility-longevity tradeoff function, this tradeoff function does not have any underpinning empirical data. However, we find that the form of this function must be strict; we refer the reader to Section \ref{sensitivity_male} where we provide justification for the form of Eq.~(\ref{malecomp}) and also exploration of different forms in a sensitivity analysis.}

\subsection{Grandmothers}
\label{GMsect}
We assume only post-fertile females (i.e. with ages $a\in[\tau_2(M),\tau_3]$) are eligible to grandmother, although beneficial grandmothering is known to come from younger grandmothers as well (see Sear et al. \cite{121}, Lahdenpera et al. \cite{122}). Furthermore, we assume grandmothers only subsidize their daughters' fertility. We let $G(a,M,L,t)$ denote the density of females eligible to grandmother and in their fertile period have given birth to females who have reached age $a$ at time $t$ and have life-history traits $M$ and $L$. Then $\frac{G(a,M,L,t)}{u_f(a,M,L,t)}$ gives the ratio of fertile females with a living mother able to support them.  We assume the sole benefit of grandmothers is shortening the birth intervals of their daughters by a factor of $m$ and we express this in the model by multiplying the birth rate $b$ by $B\left( \frac{G(a,M,L,t)}{u(a,M,L,t)} \right)$, where $B(x)$ is the benefit to net growth from grandmothering effects, given by

\begin{equation}
\label{benefit}
B(x) = m x + (1-x).
\end{equation}
\\
In Table \ref{lifehistoryparatable} we choose $m=3.0$ (which corresponds to a birth rate of 0.33 births/year when a grandmother is present) by assuming the age of weaning to be 2 years (based on Sear \& Mace \cite{167} who find that children are heavily dependent on their mothers before the age of 2 years) and that grandmothers provision weaned grandchildren, which allows their daughters to give birth to next offspring as soon as their current offspring are weaned. Using the assumption of a 1 year period for both conception and gestation, this gives a mean interbirth interval of approximately 3 years or a mean birth rate of 0.33 births/year.
\subsection{{Boundary condition}}
\label{bcsect}

According to the assumptions stated in Sections \ref{fertsect}-\ref{GMsect}, we define the boundary condition to system (\ref{mckendrickpdegm}) to be

\begin{equation}
\label{boundary}
\begin{aligned}
u_m(0,M,L,t) &= u_f(0,M,L,t) = \\
&\frac{1}{2 (1+\Phi(t))} \int_0^1 \int_0^1 S_f (y_1,y_2,t) \int_0^1\int_0^1 \frac{\phi(x_2) S_m (x_1,x_2,t)}{\bar{S}_m (t)} N \left( \frac{y_1 + x_1}{2},\frac{y_2 + x_2}{2},M,L \right) \, dx_1 dx_2 \, dy_1 \, dy_2
\end{aligned}
\end{equation}
\\
where $\bar{S}_m(t) = \int_0^1 \int_0^1 \phi(L) \int_{15}^{\tau_3} u_m(a,M,L,t) \, da \, dM \, dL$, $S_m (M,L,t) = \int_{15}^{\tau_3} u_m(a,M,L,t) \, da$ and

\begin{equation}
\label{comp}
S_f (M,L,t) = \int_{\tau_1(L)}^{\tau_2(M)} b(M,L) B\left( \frac{G(a,M,L,t)}{u_f(a,M,L,t)} \right) u_f(a,M,L,t) \, da.
\end{equation}
\\
{We note that we define Eq.~(\ref{comp}) only for $\tau_2(M) > \tau_1(L)$, otherwise it is equal to 0.}
\\
\\
The function $S_f(y_1,y_2,t)$ gives the number of offspring born by fertile females with traits $M=y_1$ and $L=y_2$ at time $t$, accounting for help from grandmothers. The expression $\frac{\phi(x_2) S_m (x_1,x_2,t)}{\bar{S}_m (t)}$ in the inner integral of Eq.~(\ref{boundary}) is the normalised distribution in $M$ and $L$ of males, accounting for male competition. It represents the probability of a female mating with a male with traits $M=x_1$ and $L=x_2$. The function $N(x,y,M,L)$ is the mutation kernel, given by

\begin{equation}
\label{mutationkernel}
N(x,y,M,L) = 
\begin{cases}
\begin{aligned}
(1-p)\delta(M-x)\delta(L-y) \quad\quad &\text{ if } \quad\quad M=x \text{ and } L=y,\\
\frac{p}{(2h)^2} \quad\quad &\text{ if } \quad\quad h \geq |M-x| \text{ and } h \geq |L-y|,\\
0 \quad\quad &\text{otherwise}.
\end{aligned}
\end{cases}
\end{equation}
\\
\begin{figure}[H]
  \centerline{\includegraphics[scale=0.4]{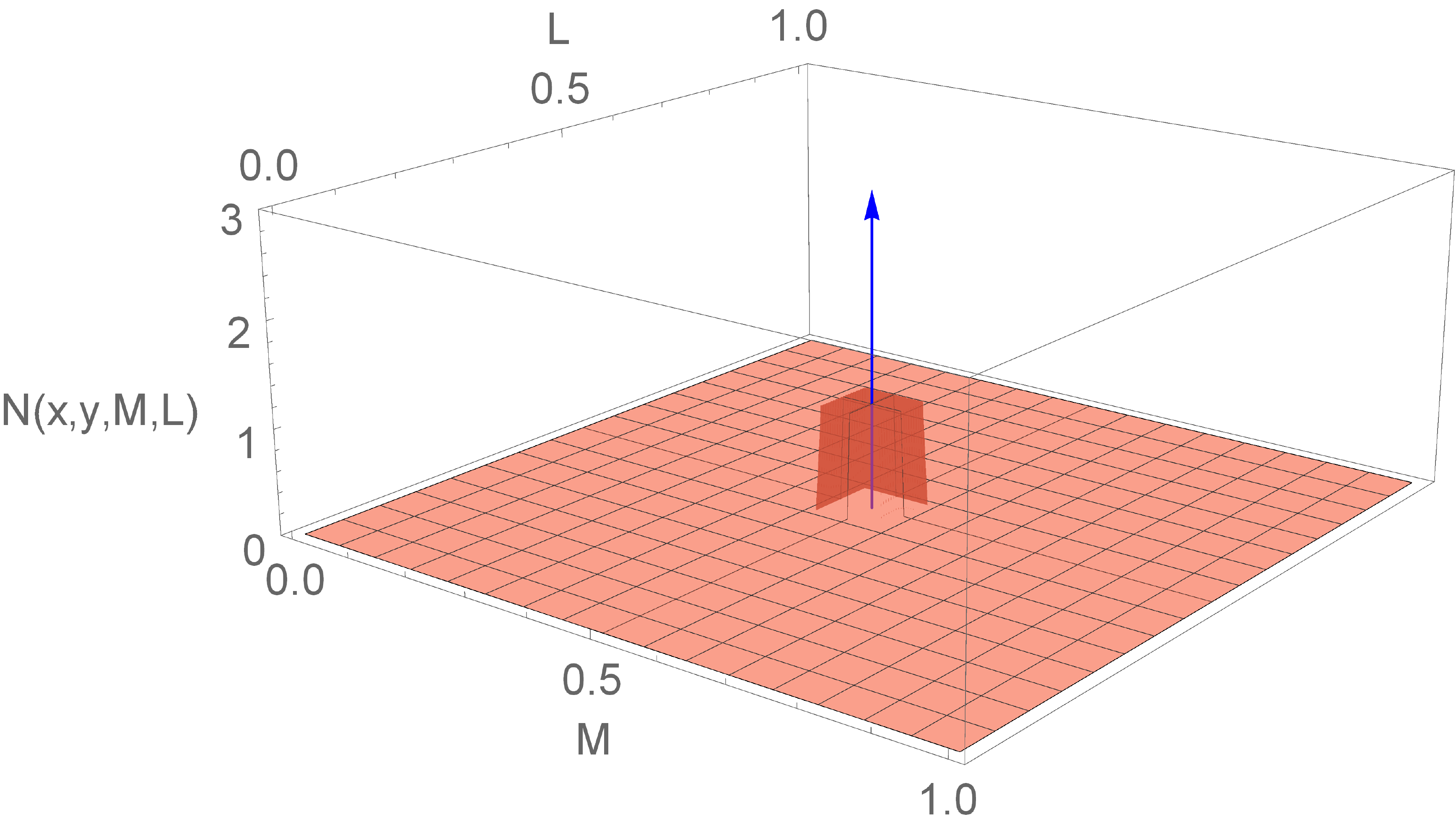}}
  \caption{Plot of the mutation kernel $N(x,y,M,L)$ with $x=y=0.5$. The blue arrow represents $(1-p)\delta(M-x)\delta(L-y)$ and the box surrounding it has height $\frac{p}{(2h)^2}$ and width $h$.}
  \label{mutationkernelfigure}
\end{figure}
The expression $N \left( \frac{y_1 + x_1}{2},\frac{y_2 + x_2}{2},M,L \right)$ in Eq.~(\ref{boundary}) is used to obtain the mean values of the traits $M$ and $L$ of the parents, with a probability of $1-p$ to not mutate and a probability of $p$ to mutate a maximum distance of $h$ (we set $h=0.05$ and $p=0.05$, as described in Section \ref{matingoffsect}). Thus, $S_f (y_1,y_2,t) \frac{\phi(x_2) S_m (x_1,x_2,t)}{\bar{S}_m (t)} N \left( \frac{y_1 + x_1}{2},\frac{y_2 + x_2}{2},M,L \right)$ gives the density of offspring with longevity $L$ and age at last birth $M$ produced by fertile females and males with $(M,L) = (y_1,y_2)$ and $(x_1,x_2)$ respectively. Integrating over $y_1$, $y_2$, $x_1$ and $x_2$, as shown in Eq.~(\ref{boundary}), gives the total density of offspring with longevity $L$ and age at last birth $M$ born at time $t$. As we have assumed an equal sex ratio, the total density of offspring is divided by 2 to find the total density of male and female offspring entering the system. Finally, the factor $\frac{1}{1+\Phi(t)}$ in Eq.~(\ref{boundary}) is to provide a density dependent form of growth for the population by regulating the total number of offspring born via the total number of females in the system, as described in Section \ref{matingoffsect}.
\\
\\
The function $G(a,M,L,t)$ in Eq.~(\ref{comp}) gives the density of mothers, who have transitioned into grandmothering, of females with age $a$ at time $t$, age at last birth $M$ and longevity $L$. It is computed via

\begin{equation}
\label{GMcomp}
\begin{aligned}
G(a,M,L,t) = \frac{\exp \left( -\int_{0}^{a} \mu(s,L) \, ds \right)}{2\bar{S}_m(t-a)(1+\Phi(t-a))} \int_0^1 \int_0^1 & \bar{S}_f (a,y_1,y_2,t-a)  \\
&\int_0^1\int_0^1 \phi(x_2) S_m (x_1,x_2,t-a) N \left( \frac{y_1 + x_1}{2},\frac{y_2 + x_2}{2},M,L \right) \, dx_1 \, dx_2 \, dy_1 \, dy_2,
\end{aligned}
\end{equation}
\\
where 

\begin{equation}
\label{Sfbar}
\begin{aligned}
\bar{S}_f (a,M,L,t-a) = b(M,L) \int_{\tau_1(L)}^{\tau_2(M)}& \mathds{1}_A(\bar{a}) \exp \left( -\int_{\bar{a}}^{\bar{a}+a} \mu(s,L) \, ds \right) \\
&B\left( \frac{G(\bar{a},M,L,t-a)}{u_f(\bar{a},M,L,t-a)} \right) u_f(\bar{a},M,L,t-a) \, d\bar{a},
\end{aligned}
\end{equation}
\\
and $\mathds{1}_A(x)$ is the indicator function of the subset $A$, which is defined to be $A = [\max(\tau_2(M)-a,\tau_1(L)), \min(\tau_2(M),\tau_3-a)]$. {Similar to Eq.~(\ref{comp}), we define Eq.~(\ref{Sfbar}) only for $\tau_2(M) > \tau_1(L)$, otherwise it is equal to 0.}
\\
\\
The expression in Eq.~(\ref{Sfbar}) represents the density of surviving mothers of fertile females with age $a$ at time $t-a$ and traits $M$ and $L$. The subset $A$ of the indicator function ensures that only females with ages between $\max (\tau_2(M)-a,\tau_1(L))$ to $\min(\tau_2(M),\tau_3-a)$ are integrated over, i.e. females who were fertile at time $t-a$ but have transitioned into grandmothering (i.e. with age $a \in [\tau_2(M),\tau_3]$) at time $t$. The factor $\exp \left( -\int_{\bar{a}}^{\bar{a}+a} \mu(x,L) \, dx \right)$ in the integrand of Eq.~(\ref{Sfbar}) gives the proportion of mothers surviving from time $t-a$ to time $a$, given that they have longevity $L$. Additionally, the factor $\exp \left( -\int_{0}^{a} \mu(x,L) \, dx \right)$ in Eq.~(\ref{GMcomp}) gives the proportion of daughters surviving from birth to time $t$. The combination of these two factors gives the expected proportion of the mother and daughter pairs surviving from time $t-a$ to time $t$. As we have assumed that grandmothers take care of all their daughters, providing that the grandmother is alive but not frail, individuals who have a birth rate of $b(M,L)$ will count as $b(M,L)/2$ grandmothers at every time $t$. Thus, a grandmother with life-history traits $M$ and $L$ takes care of 1 daughter every $2/b(M,L)$ years.

\begin{table}[H]
  \centering
  \caption{Parameter values}
    \begin{tabular}{lll}
    \toprule
    Symbol & Definition & Value\\
    \midrule
    $m$  & Benefit received by fertile female from grandmothering & $3.0$  \\
    $\tau_1$	& Age of female sexual maturity & Function of $L$, see Eq.~(\ref{tau1}) \\
    $\tau_2$	& Age female fertility ends & Function of $M$, see Eq.~(\ref{tau2}) \\
    $\tau_3$ & Age of frailty & $75$ \\
    $\rho$ & Parameter controlling male competition with respect to $L$, see Eq.~(\ref{malecomp})& $8$ \\
    $\mu_j$ & Early mortality rate & $0.1$ \\
    $\mu_a$ & Middle mortality rate & $0.06$ \\
    $\mu_s$ & Late mortality rate & $0.08$ \\
    \bottomrule
    \end{tabular}
  \label{lifehistoryparatable}
    \\[5pt]
\end{table}

\section{Results}

\subsection{Time evolution without grandmothering}

\begin{figure}[H]
  \hspace{1mm}\centerline{
  \subfloat[]{\label{chimps_Levol}\includegraphics[width=0.45\textwidth]{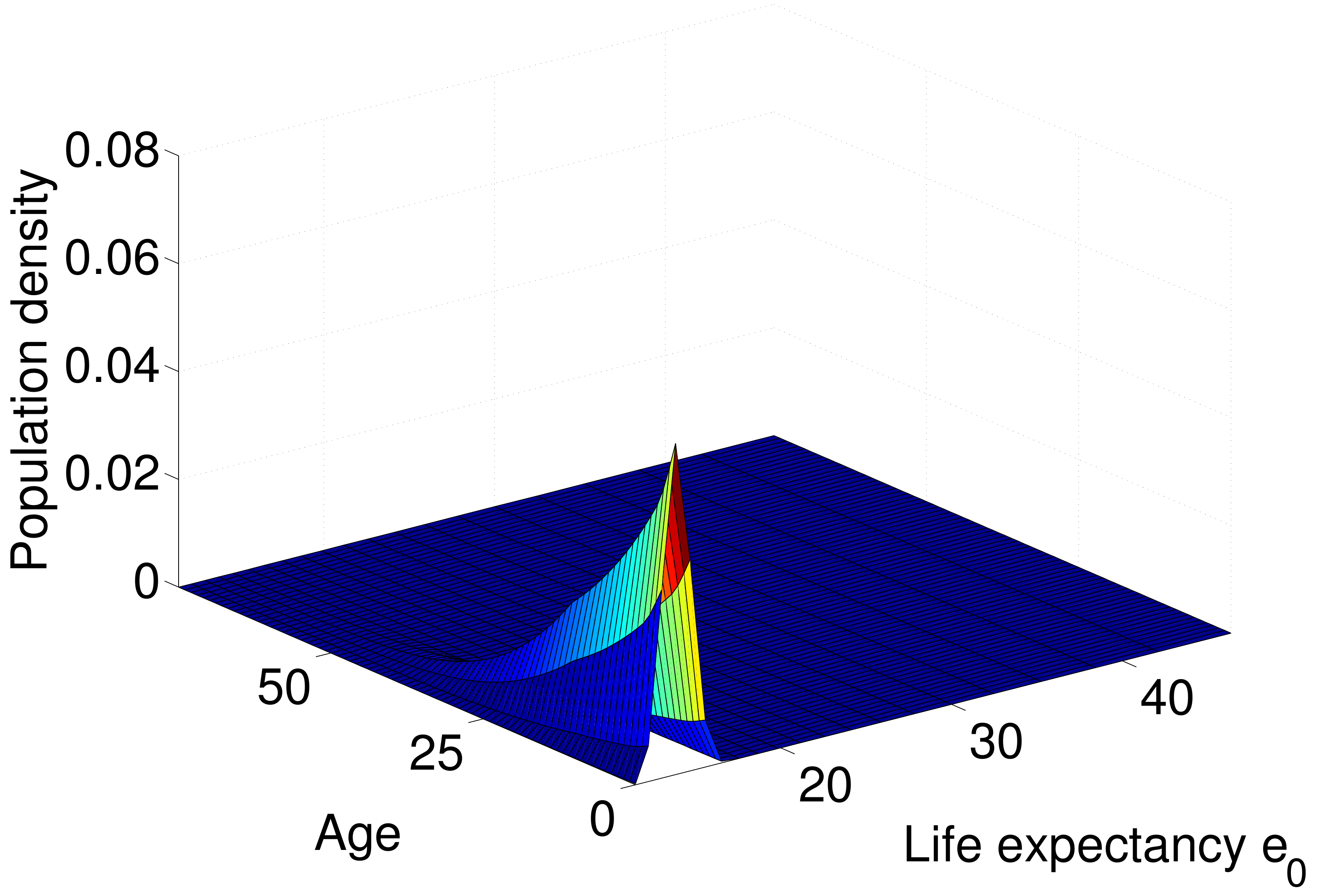}}
  \hspace{10mm}
  \subfloat[]{\label{chimps_Mevol}\includegraphics[width=0.45\textwidth]{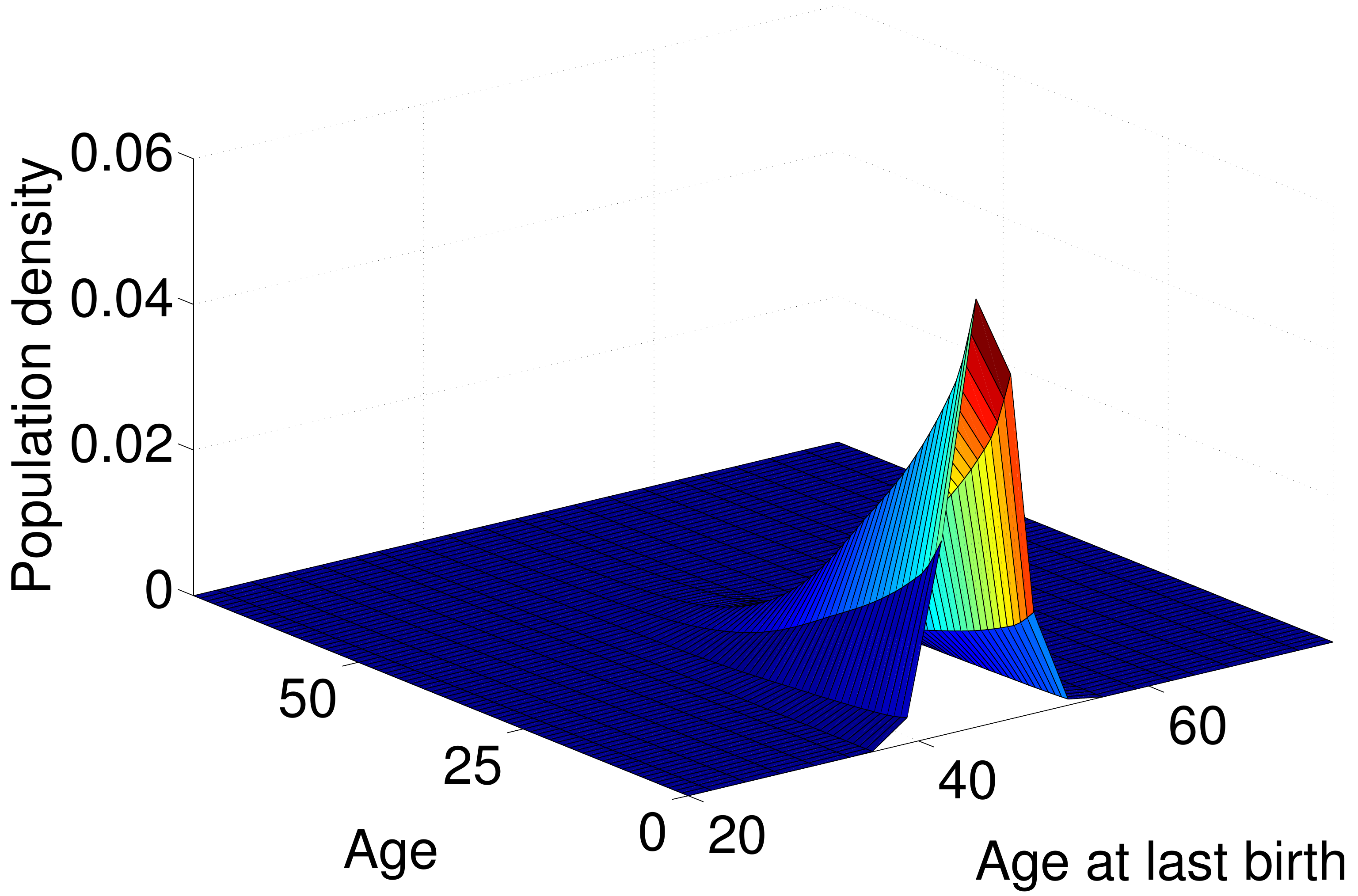}}}
  \caption{Plots of the system without grandmothering at equilibrium. Plots (a) and (b) show $\int_0^1 u_f \, dM$ and $\int_0^1 u_f \, dL$ respectively.}
  \label{plots_chimps}
\end{figure}

\begin{figure}[H]
  \centerline{
  \includegraphics[width=0.45\textwidth]{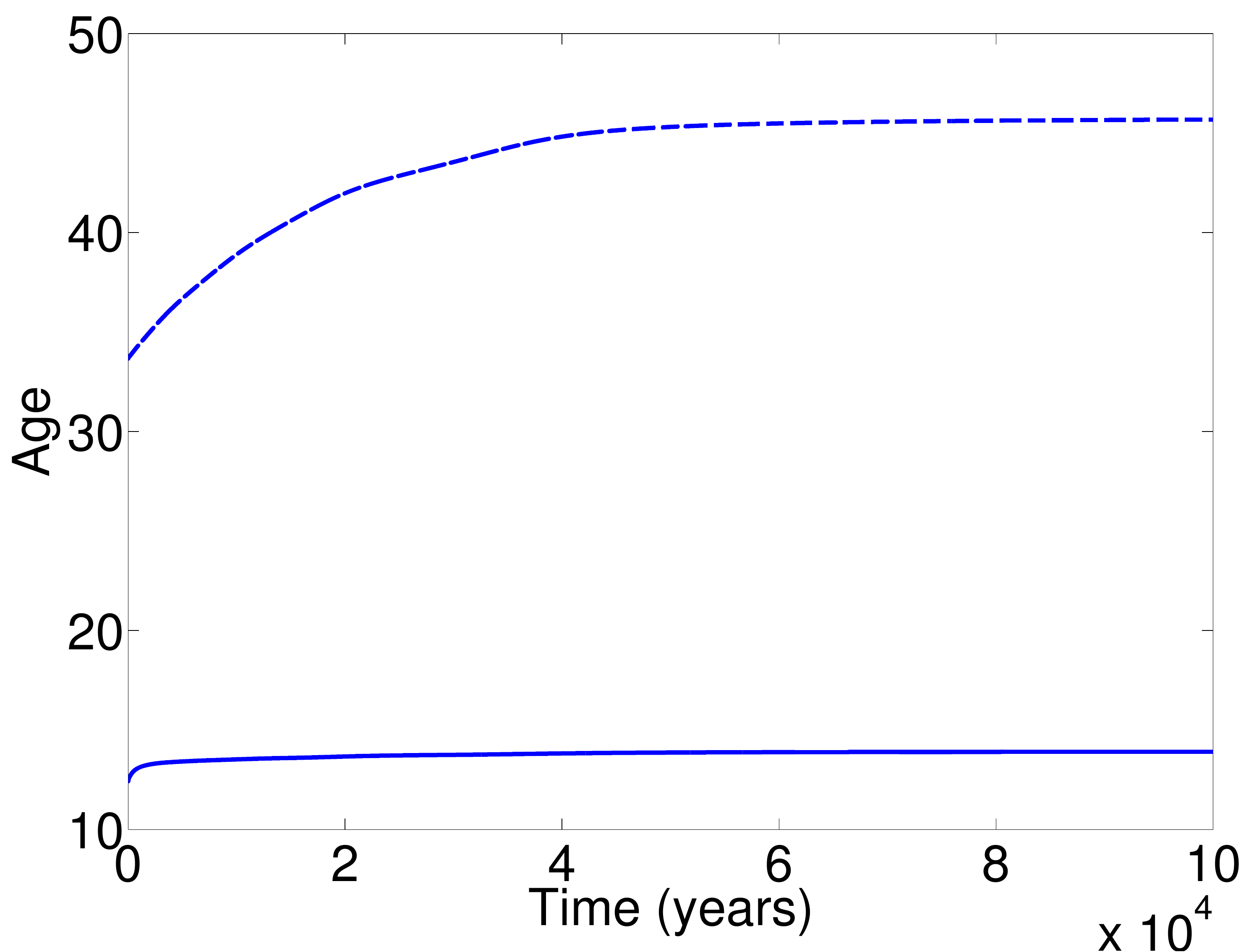}}
  \caption{Time evolution of the average adult lifespan (solid line) and average age at last birth (dashed line), without grandmothering.}
  \label{plots_chimps_evol}
\end{figure}

For the case without grandmothering, we set $B(x) = 1$ instead of its definition in Eq.~(\ref{benefit}). We use the initial condition

\begin{equation}
u_f(a,M,L,0) = u_m(a,M,L,0) =
\begin{cases}
1 \quad\quad \text{if } \quad\quad a \leq 5, 0.2 \leq M \leq 0.3, 0 \leq L \leq 0.1,\\
0 \quad\quad \text{otherwise.}
\end{cases}
\end{equation}
\\
However, any initial condition close to $L=0.2$ and $M=0.5$ will result in the population converging to the equilibrium shown in Figure \ref{plots_chimps}, with an average adult lifespan of 13.92 years and average age at last birth of 45.76 years.
\\
\\
Due to the initial condition being close to the equilibrium value for average adult lifespan, we can see from Figure \ref{plots_chimps_evol} that the average adult lifespan converges relatively quickly compared to the average age at last birth. The speed at which the average age at last birth converges to the equilibrium decreases as the population evolves towards the equilibrium; this is due to few individuals surviving to ages 40 and beyond (observable from the tails of the solution in Figure \ref{plots_chimps}) at $L=0.2$, which leads to weak selection for an age at last birth of 40 and above.
\\
\\
Note that the equilibrium we show in Figure \ref{plots_chimps} is only locally stable. For initial conditions with $L$ values significantly above $L=0.2$, the population diverges from this equilibrium; the average adult lifespan increases to 34 years, but the population goes extinct as the net reproductive rate is negative at such lifespans without grandmothering. Without grandmothering, the only equilibrium we observed with a net reproductive rate greater than 0 is the one shown in Figures \ref{plots_chimps}-\ref{plots_chimps_evol}.

\subsection{Time evolution with grandmothering}

\begin{figure}[H]
  \hspace{1mm}\centerline{
  \subfloat[]{\label{GM_Levol}\includegraphics[width=0.45\textwidth]{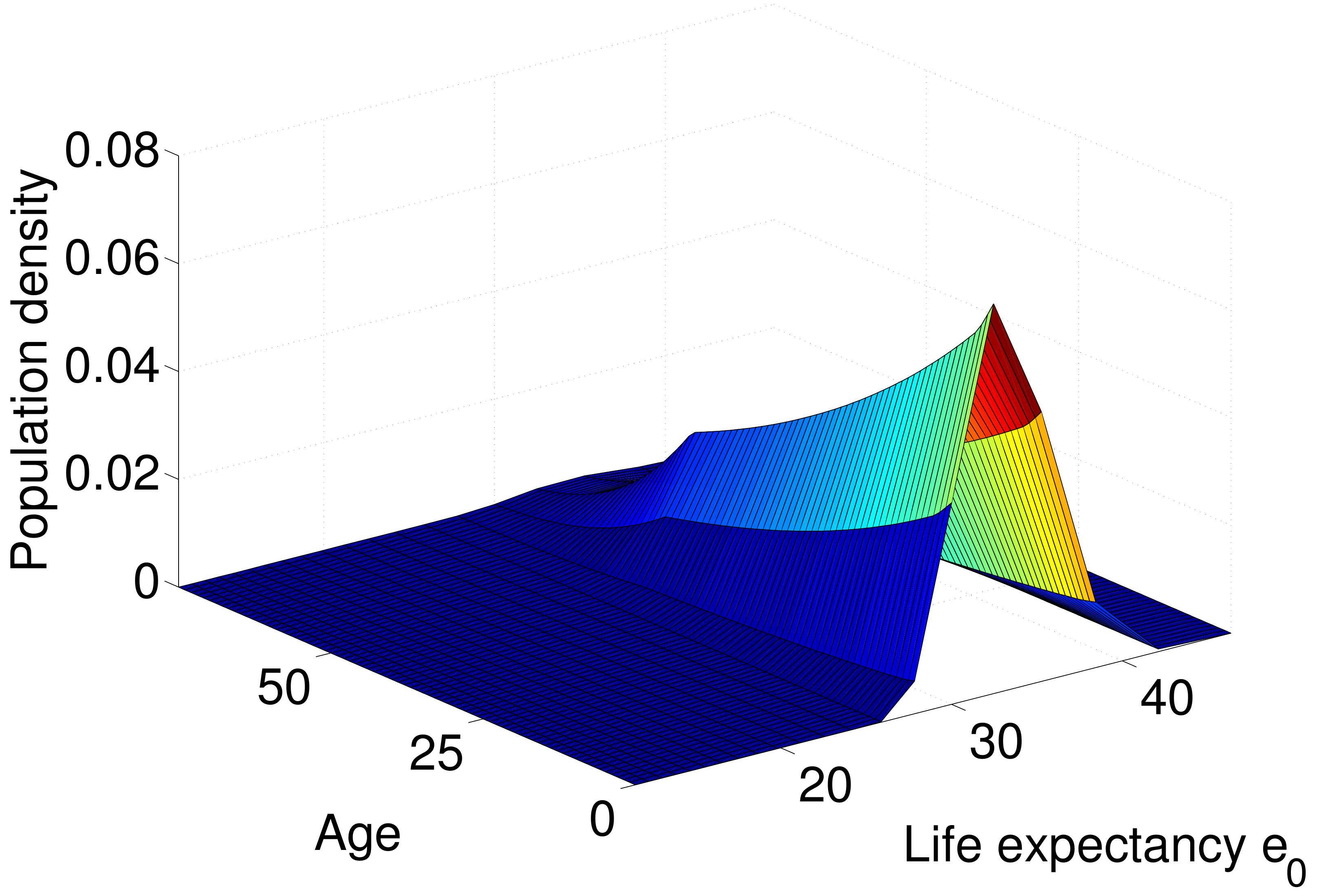}}
  \hspace{10mm}
  \subfloat[]{\label{GM_Mevol}\includegraphics[width=0.45\textwidth]{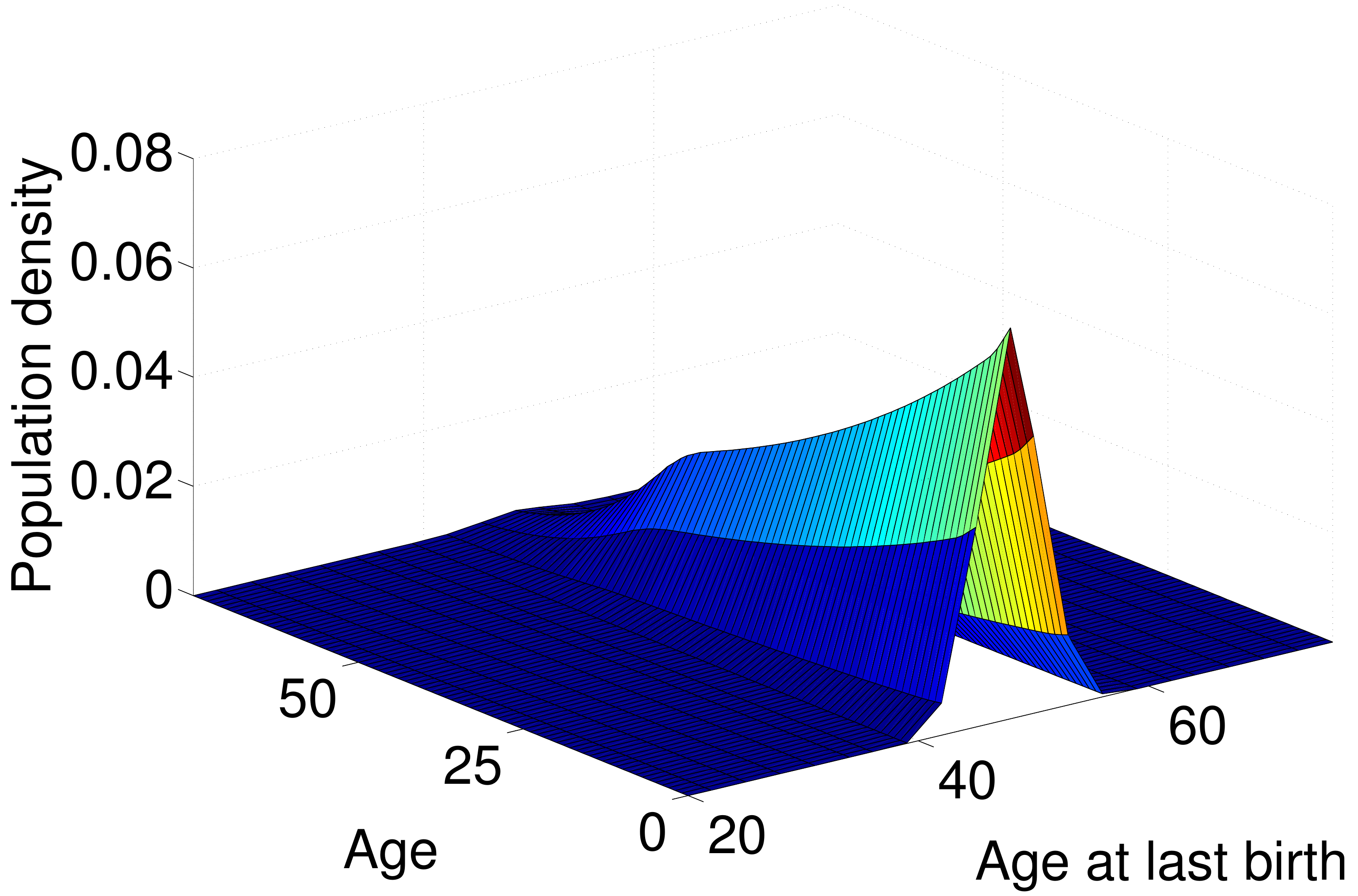}}}
  \caption{Plots of the system with grandmothering at equilibrium. Plots (a) and (b) show $\int_0^1 u_f \, dM$ and $\int_0^1 u_f \, dL$ respectively.}
  \label{plots_GM}
\end{figure}

\begin{figure}[H]
  \centerline{
  \includegraphics[width=0.45\textwidth]{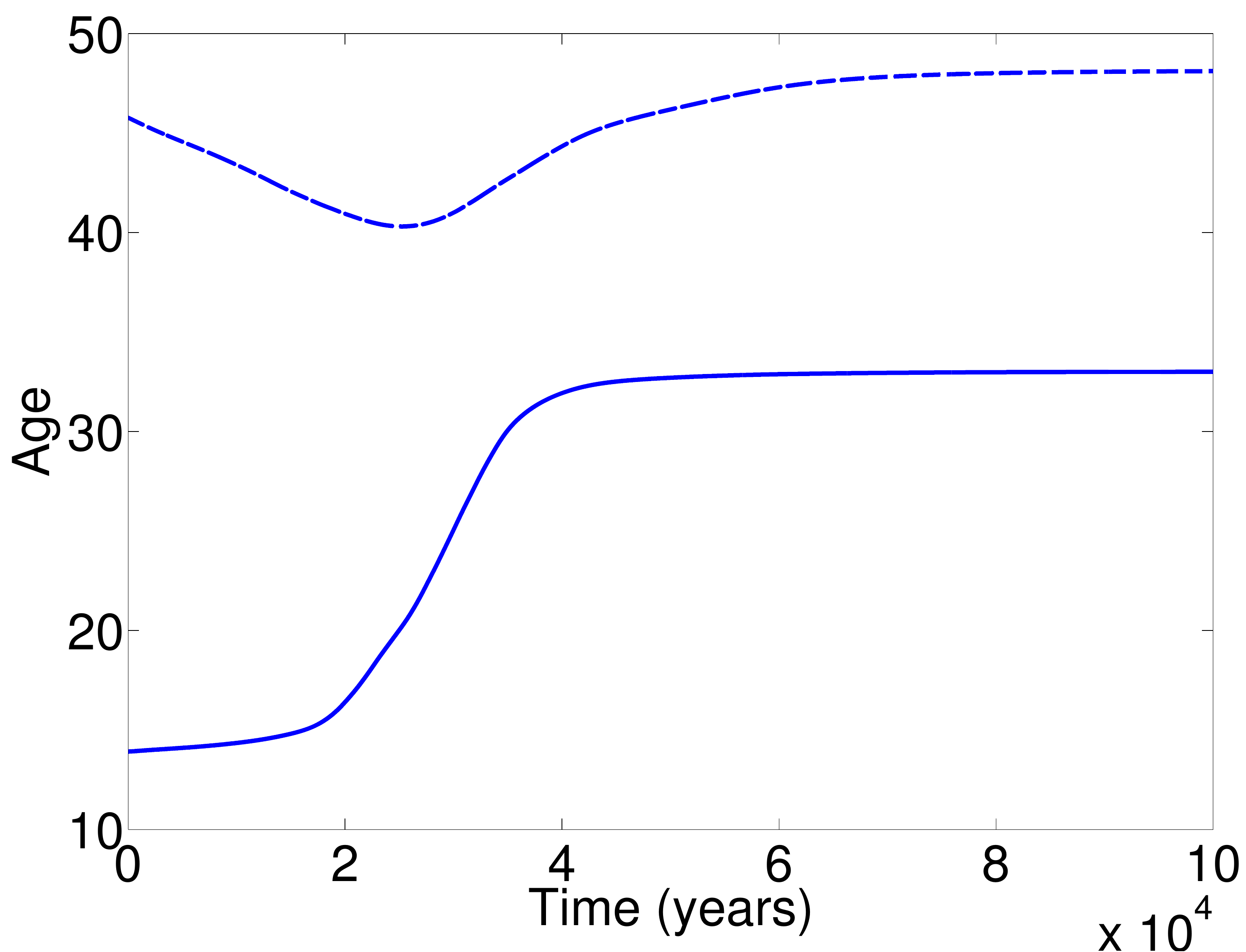}}
  \caption{Time evolution of the average adult lifespan (solid line) and average age at last birth (dashed line), with grandmothering.}
  \label{plots_GM_evol}
\end{figure}

Using the equilibrium from Figure \ref{plots_chimps}, we let post-fertile females grandmother, i.e. help their daughters by increasing their birth rate by $B(x)$, see Eq.~(\ref{benefit}). From Figure \ref{plots_GM_evol}, we see that it takes roughly 30,000 years for the population to progress from great ape-like longevities to hunter-gatherer longevities, while keeping the average age at last birth at similar levels. At the hunter-gatherer equilibrium shown in Figure \ref{plots_GM_evol}, the average adult lifespan is 33.01 years and the average age at last birth is 48.14 years.
\\
\\
Noteworthy is the decline in average age at last birth early in the evolutionary trajectory, which increases again when the average adult lifespan rises past a certain threshold as seen in Figure \ref{plots_GM_evol}. When the adult lifespan is low and the age at last birth decreases, the benefit of having more grandmothers in the population heavily outweighs the decrease in fertile females. This results in selection for a lower age at last birth to increase the number of grandmothers in the population. Moreover, at life expectancies close to the great ape-like equilibrium values, the proportion of females that are post-fertile is very low, which contributes to the sharp decline in age at last birth during the earlier phases of the evolutionary trajectory.

\section{Sensitivity analyses}
\label{sensitivitysect}
\subsection{Male fertility-longevity tradeoff}
\label{sensitivity_male}
\begin{figure}[H]
  \centerline{
  \includegraphics[width=0.45\textwidth]{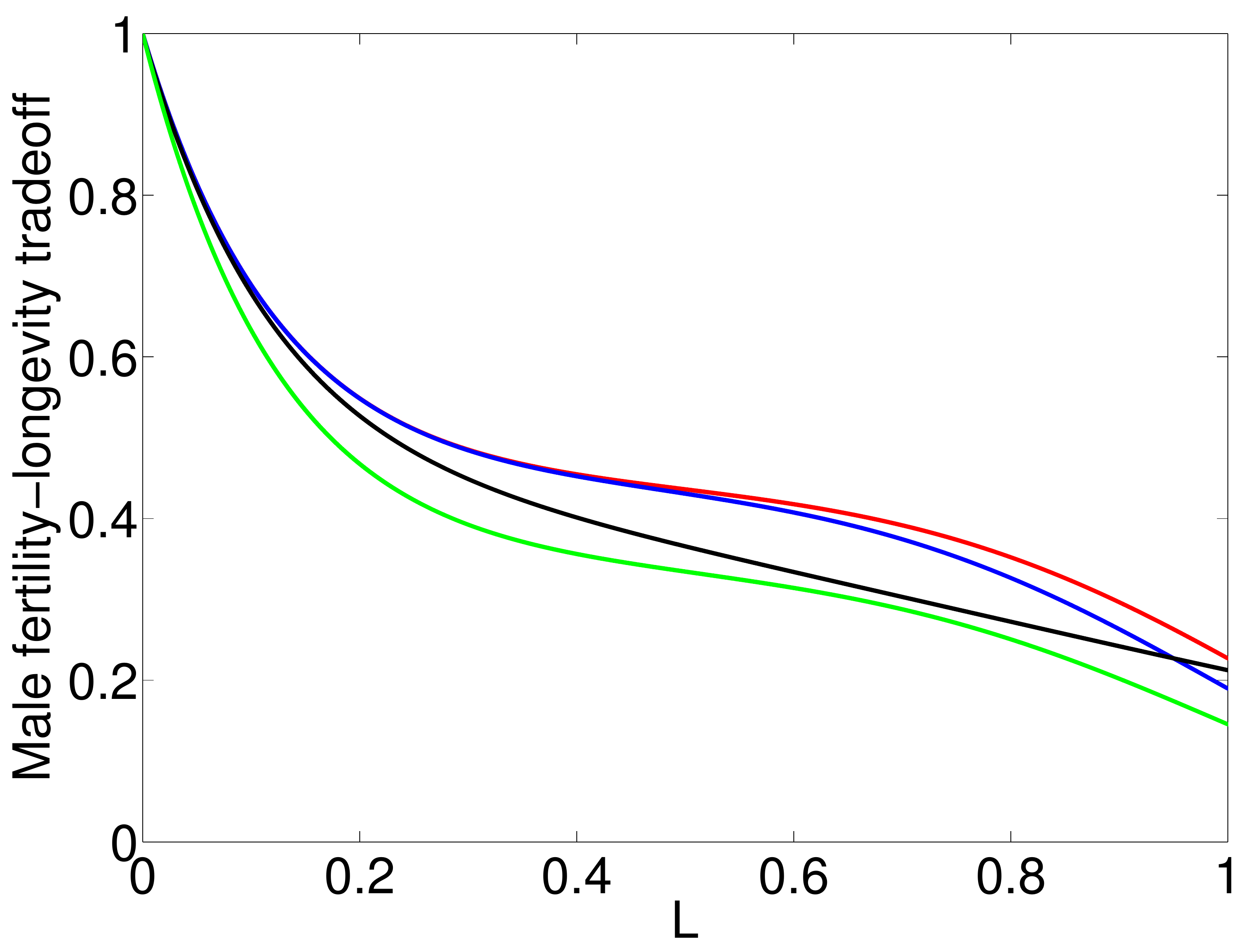}}
  \caption{Male fertility-longevity tradeoff functions $\phi(L)$ (blue), $\phi_1(L)$ (red), $\phi_2(L)$ (black) and $\phi_3(L)$ (green).}
  \label{sensitivity_matingfail}
\end{figure}

We examine the equlibria of the system for different, but qualitatively similar, male fertility-longevity tradeoff functions, $\phi_1(L)$, $\phi_2(L)$ and $\phi_3(L)$, where 

\begin{equation}
\begin{aligned}
\phi_1(L) &= \left(2.3 e^{-\rho L}+1\right)\exp\left( - 0.65 L^{5} \right)/2.3,\\
\phi_2(L) &= \left(2.3 e^{-\rho L}+1\right)\exp\left( - 0.65 (L+0.05)^{2} \right)/2.3,\\
\phi_3(L) &= \left(3 e^{-\rho L}+1\right)\exp\left( - 0.65 (L+0.05)^{5} \right)/3.
\end{aligned}
\end{equation}
\\
See Figure \ref{sensitivity_matingfail} for a comparison of these functions and Eq.~({\ref{malecomp}}).

\begin{figure}[H]
  \hspace{1mm}\centerline{
  \subfloat[]{\label{sensitivity_phi_chimps_M}\includegraphics[width=0.45\textwidth]{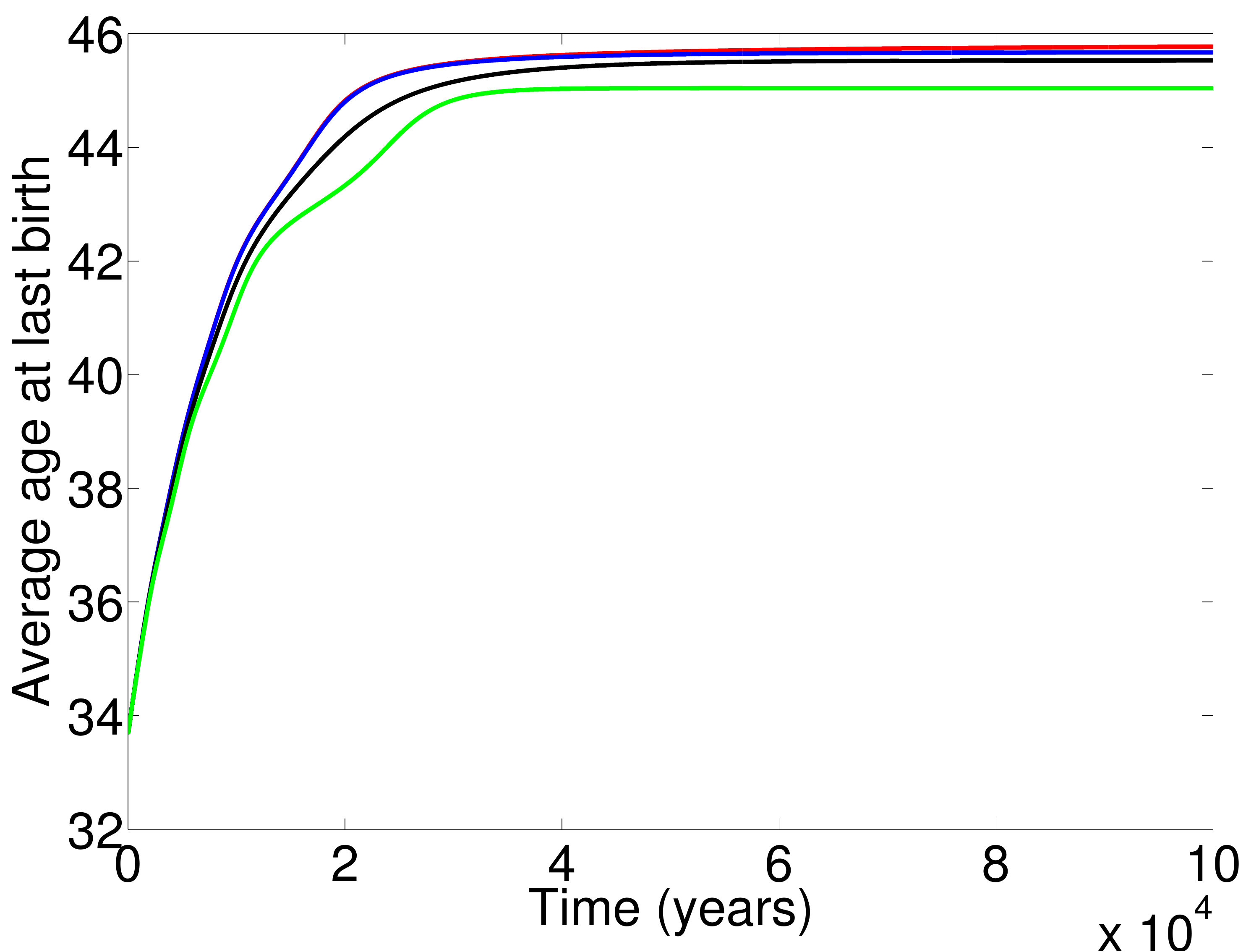}}
  \hspace{5mm}
  \subfloat[]{\label{sensitivity_phi_chimps_L}\includegraphics[width=0.45\textwidth]{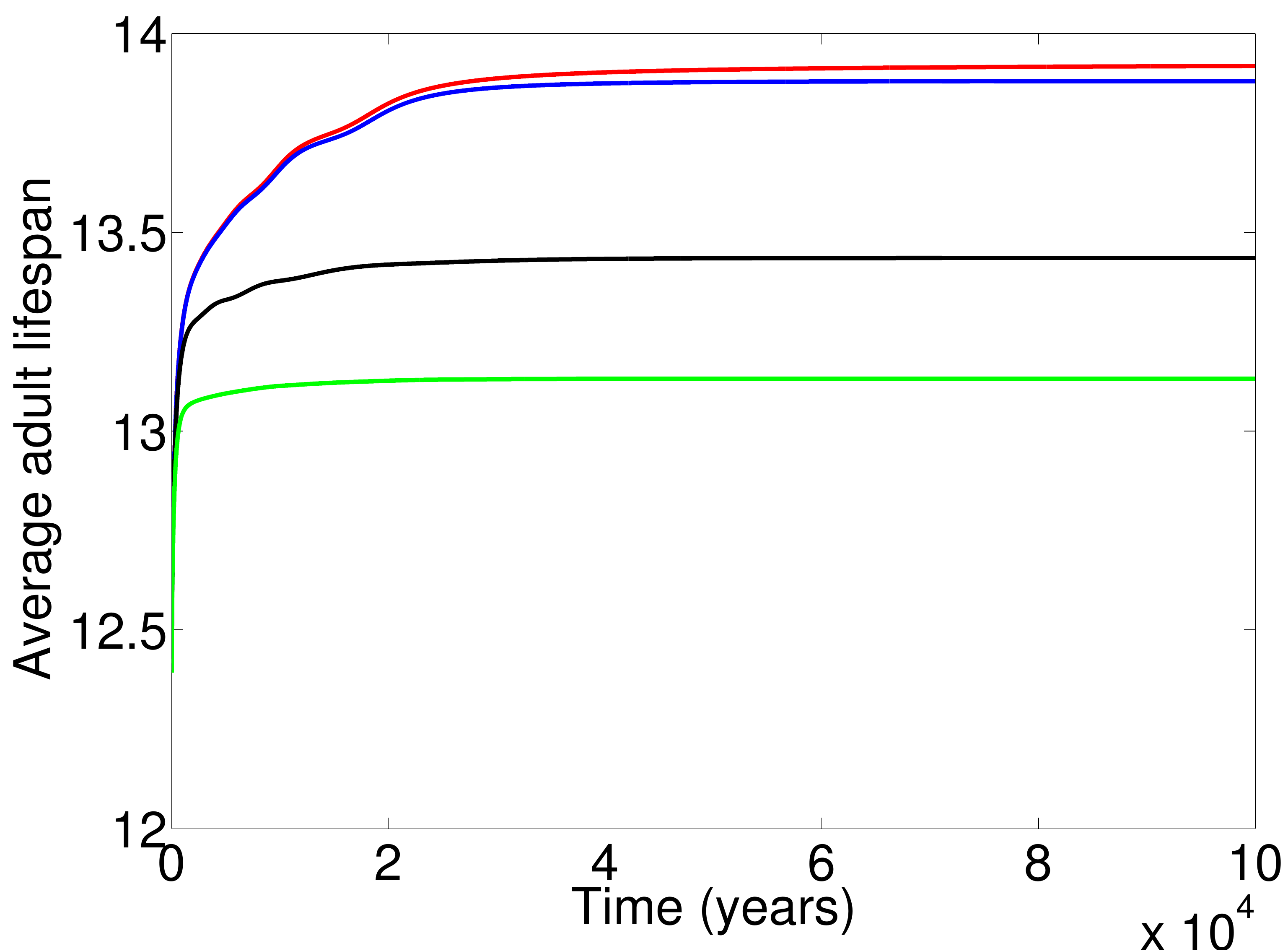}}}
  \caption{Time evolutions of the system without grandmothering with $\phi(L)$ (blue), $\phi_1(L)$ (red), $\phi_2(L)$ (black) and $\phi_3(L)$ (green).}
  \label{sensitivity_phi_chimps}
\end{figure}

\begin{figure}[H]
  \hspace{1mm}\centerline{
  \subfloat[]{\label{sensitivity_phi_GMs_M}\includegraphics[width=0.45\textwidth]{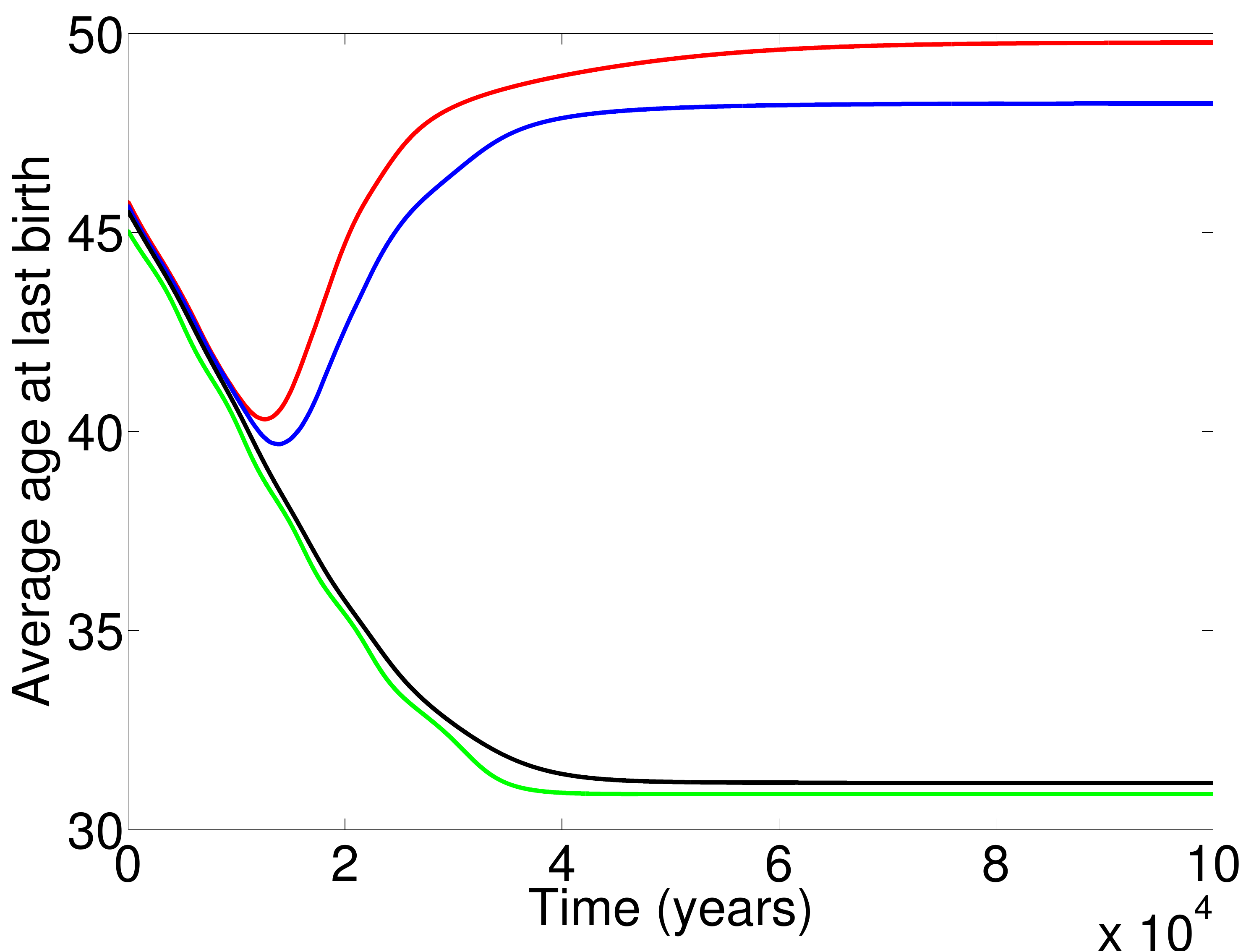}}
  \hspace{5mm}
  \subfloat[]{\label{sensitivity_phi_GMs_L}\includegraphics[width=0.45\textwidth]{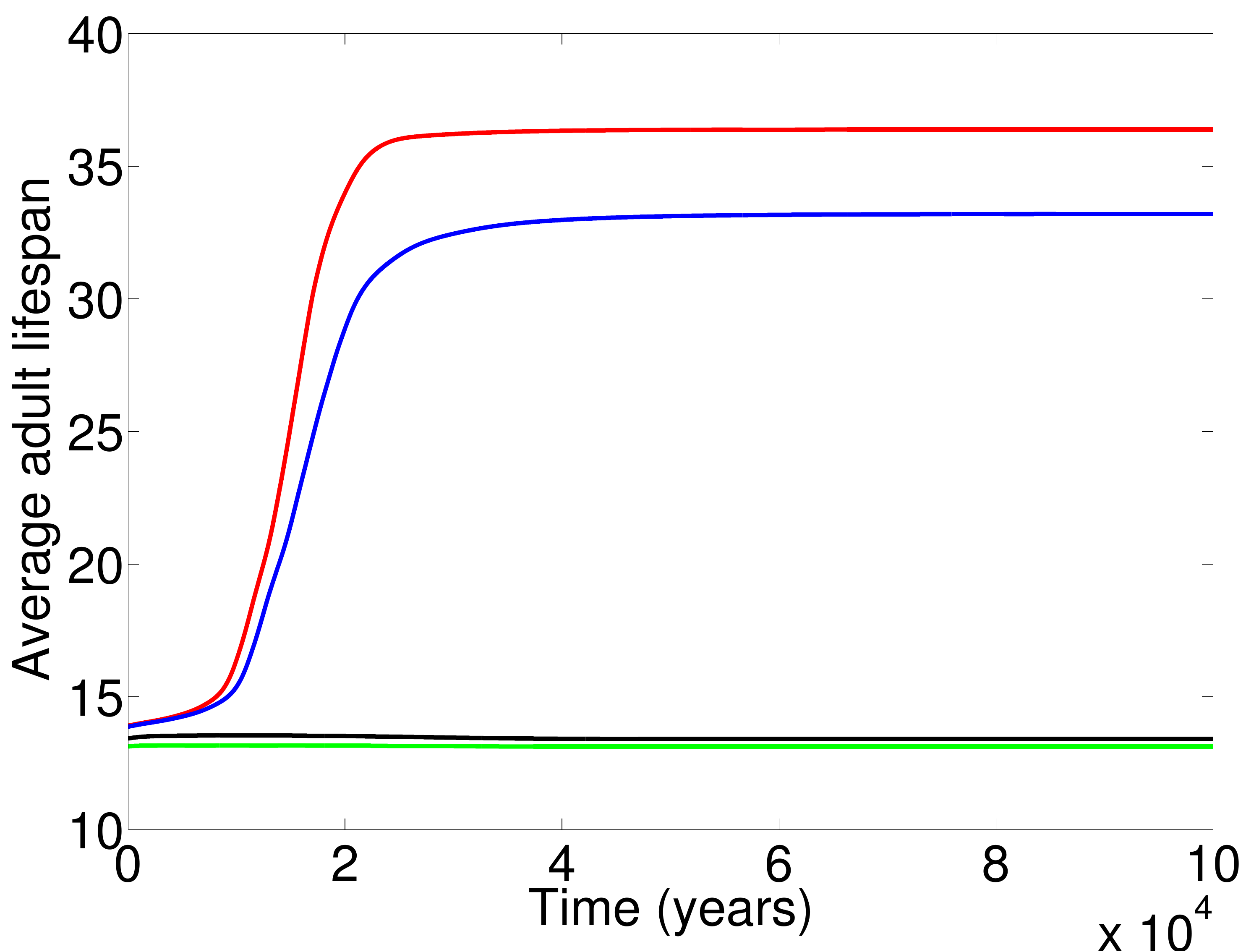}}}
  \caption{Time evolutions of the system with grandmothering with $\phi(L)$ (blue), $\phi_1(L)$ (red), $\phi_2(L)$ (black) and $\phi_3(L)$ (green). The initial condition used was the corresponding equilibria of the system without grandmothering, shown in Figure \ref{sensitivity_phi_chimps}.}
  \label{sensitivity_phi_GM}
\end{figure}

Without grandmothering, the shape of the male fertility-longevity tradeoff function has a small effect on the equilibrium value for average adult lifespan and average age at last birth, as shown in Figure \ref{sensitivity_phi_chimps}. However, small changes in the steepness of the male fertility-longevity function at low $L$ were found to be critical to adjusting the average adult lifespan of the great ape-like equilibrium; they also determined whether a basin of attraction was present and the strength of the attraction if it does exist. For male fertility-longevity functions that are too steep at low $L$, the generated populations cannot escape the great ape-like equilibria when grandmothering is introduced in the system. In Figure \ref{sensitivity_phi_GM} we show that this is the case for $\phi_2(L)$ and $\phi_3(L)$, whereas $\phi(L)$ and $\phi_1(L)$ are less steep at low $L$ and hence produce simulations where a transition to the increased longevities, while maintaining the average age at last birth, is possible.
\\
\\
Similarly to the case without grandmothering, Figure \ref{sensitivity_phi_GM} shows that the decline in the male fertility-longevity tradeoff function at high $L$ values greatly affects the average adult lifespan at the hunter-gatherer equilibrium. This is evident from a comparison of the equilibria generated by $\phi(L)$ and $\phi_1(L)$, shown in Figure \ref{sensitivity_phi_GMs_L}, where the decline in $\phi_1(L)$ occurs at higher values of $L$ and less steeply than $\phi$. Moreover, comparison of the equilibria generated by $\phi(L)$ and $\phi_1(L)$ reveal that a selection for a higher average adult lifespan is accompanied by selection for higher average age at last birth. We note that the decrease in the male fertility-longevity function for $L$ close to $1$ is necessary or else the sexual conflict from male competition pushes the average adult lifespan to values where the net reproduction rate is negative. We do not show this here due to computational limitations for the domain size. For simulations where the populations cannot escape the great ape-like equilibrium with grandmothering, the population maintains a great ape-like average adult lifespan, but lowers the average age at last birth to 31 years to maximise the net reproductive rate.

\subsection{Benefit from grandmothers}
\label{sensitivity_grandmother_sect}
\begin{figure}[H]
  \hspace{1mm}\centerline{
  \subfloat[]{\label{sensitivity_B_menopause}\includegraphics[width=0.45\textwidth]{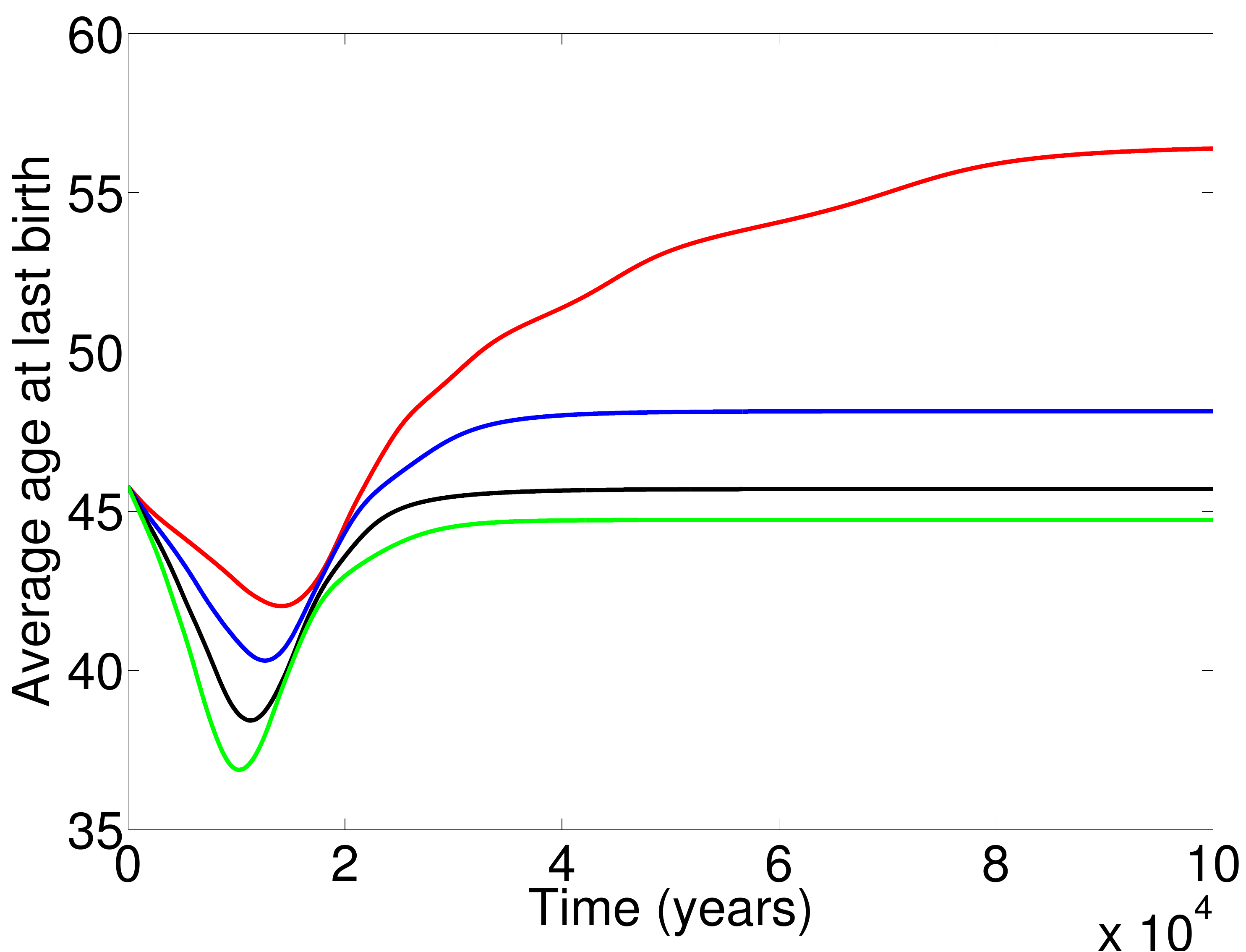}}
  \hspace{5mm}
  \subfloat[]{\label{sensitivity_B_lifespan}\includegraphics[width=0.45\textwidth]{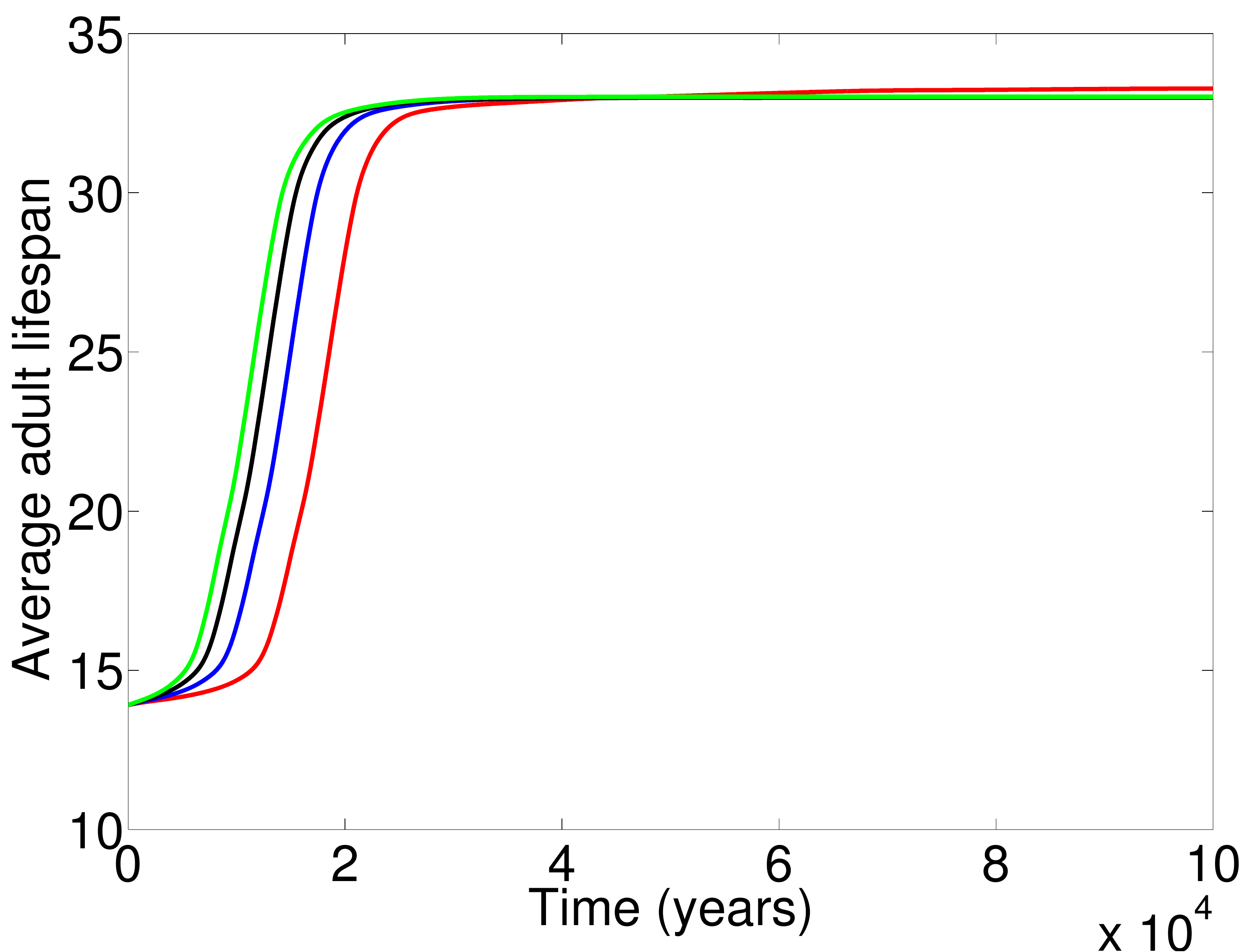}}}
  \caption{Time evolutions of the system with grandmothering with $m=2.5$ (red), $m=3.0$ (blue), $m=3.5$ (black) and $m=4.0$ (green). The initial condition used was the equilibrium of the system without grandmothering, shown in Figures \ref{plots_chimps}-\ref{plots_chimps_evol}.}
  \label{sensitivity_B}
\end{figure}

\begin{figure}[H]
  \centerline{
  \includegraphics[width=0.45\textwidth]{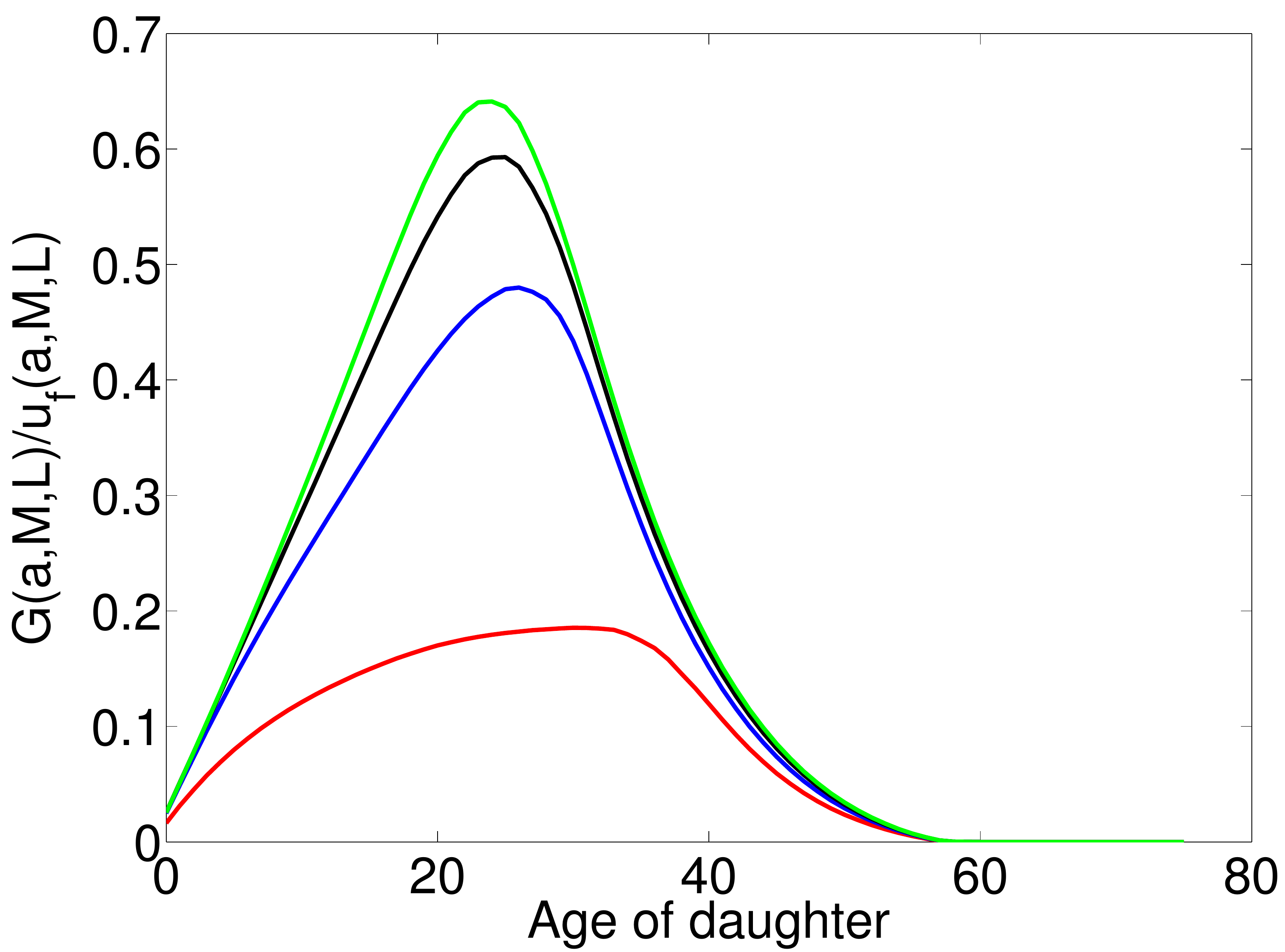}}
  \caption{The proportion of females of any age with a live mother with age $a \in [\tau_2(M),\tau_3]$ (that is, $G(a,M,L)/u_f(a,M,L)$) at the hunter-gatherer equilibrium with $m=2.5$ (red), $m=3.0$ (blue), $m=3.5$ (black) and $m=4.0$ (green).}
  \label{sensitivity_B_daughterscovered}
\end{figure}

Here we examine the trajectory of the system when the benefit to the birth rate from grandmothering ($m$ in Eq.~(\ref{benefit})) is varied. A comparison of the time evolutions in the average age at last birth and average adult lifespan with $m=2.5$, $m=3.0$, $m=3.5$ and $m=4.0$ are shown in Figures \ref{sensitivity_B_menopause}-\ref{sensitivity_B_lifespan}. These values for $m$ correspond to an average interbirth interval of 3.64, 3.33, 2.60 and 2.27 birth/year respectively. For the same variations in $m$, Figure \ref{sensitivity_B_daughterscovered} shows the proportion of daughters of any age with a live mother who has transitioned into grandmothering.
\\
\\
From Figure \ref{sensitivity_B_menopause}, the average age at last birth decreases with increasing strength of grandmothering effects, due to the average age at last birth at the equilibrium being a compromise between stopping fertility early to grandmother and continuing to be fertile to produce more offspring. This is evident from Figure \ref{sensitivity_B_daughterscovered}, which shows a positive relationship between the number of grandmothers at the equilibrium and the strength of grandmothering effects. Since having a grandmother reduces their daughter's interbirth interval by a factor of $m$, Figure \ref{sensitivity_B_menopause} implies that the average age at last birth at the hunter-gatherer equilibrium is sensitive to the reduction in the interbirth interval that a grandmother can provide. For example, $m=2.5$, corresponds to a mean interbirth interval of approximately 3.6 years, with the equilibrium value of the average age at last birth being 56 years. However, an increase to $m=3.0$ (a mean interbirth interval of approximately 3 years) drastically reduces this to 48 years.

From Figure \ref{sensitivity_B_lifespan}, differing strengths of grandmother effects have an insignificant effect on average adult lifespan. This can be surprising, since if one expects grandmothers to increase the net reproductive rate of the population, a simple conclusion is that stronger grandmother effects drive selection for longer lived grandmothers, which leads to an increase in longevity. There are two possible explanations for this to occur in the system: (1) The total increase in the net reproductive rate from the increase in grandmothers resulting from a higher life expectancy is not enough to negate the decrease in the birth rate due to the female fertility-longevity tradeoff (see Eq.~(\ref{birthrate})) and later age of sexual maturity; or (2) The skew towards the males in the adult sex ratio increases male competition, which strongly influences the equilibrium value for longevity via the male fertility-longevity tradeoff function given by Eq.~(\ref{malecomp}). Unfortunately there is no simple way of using the model presented in this study to rigorously test for possibility (1); however, the results from Section \ref{sensitivity_male} suggest that it is likely to be (2), as Figure \ref{sensitivity_phi_GM} shows that small changes in the shape of the male fertility-longevity tradeoff function result in large shifts in average adult lifespan.
\\
\\

\subsection{Age of female sexual maturity}

\begin{figure}[H]
  \hspace{1mm}\centerline{
  \subfloat[]{\label{sensitivity_tau1_menopause_chimps}\includegraphics[width=0.45\textwidth]{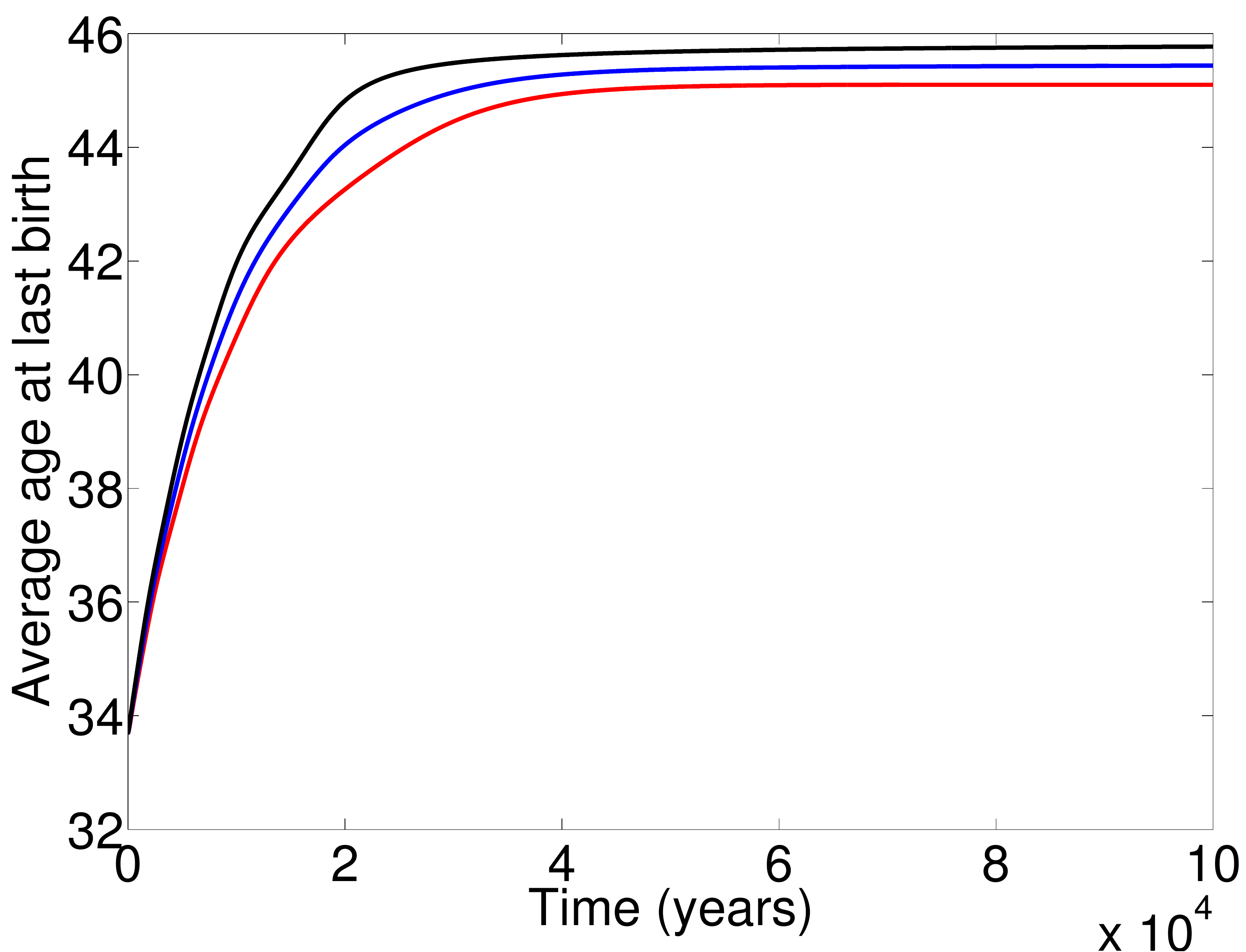}}
  \hspace{5mm}
  \subfloat[]{\label{sensitivity_tau1_e0_chimps}\includegraphics[width=0.45\textwidth]{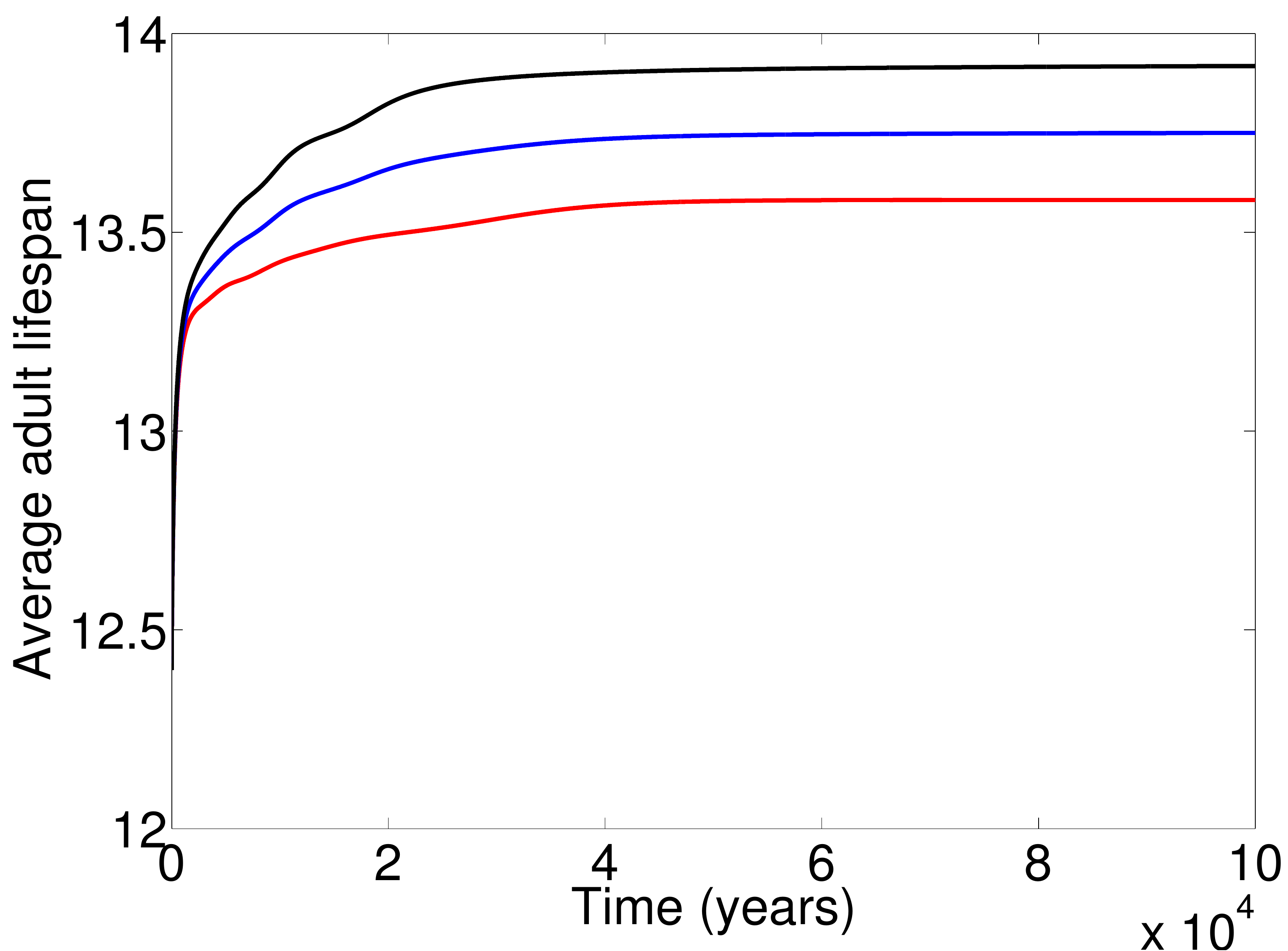}}}
  \caption{Time evolutions of the system without grandmothering with $0.8\tau_1(L)$ (red), $0.9\tau_1(L)$ (blue) and $\tau_1(L)$ (black).}
  \label{sensitivity_tau1_chimps}
\end{figure}

\begin{figure}[H]
  \hspace{1mm}\centerline{
  \subfloat[]{\label{sensitivity_tau1_menopause}\includegraphics[width=0.45\textwidth]{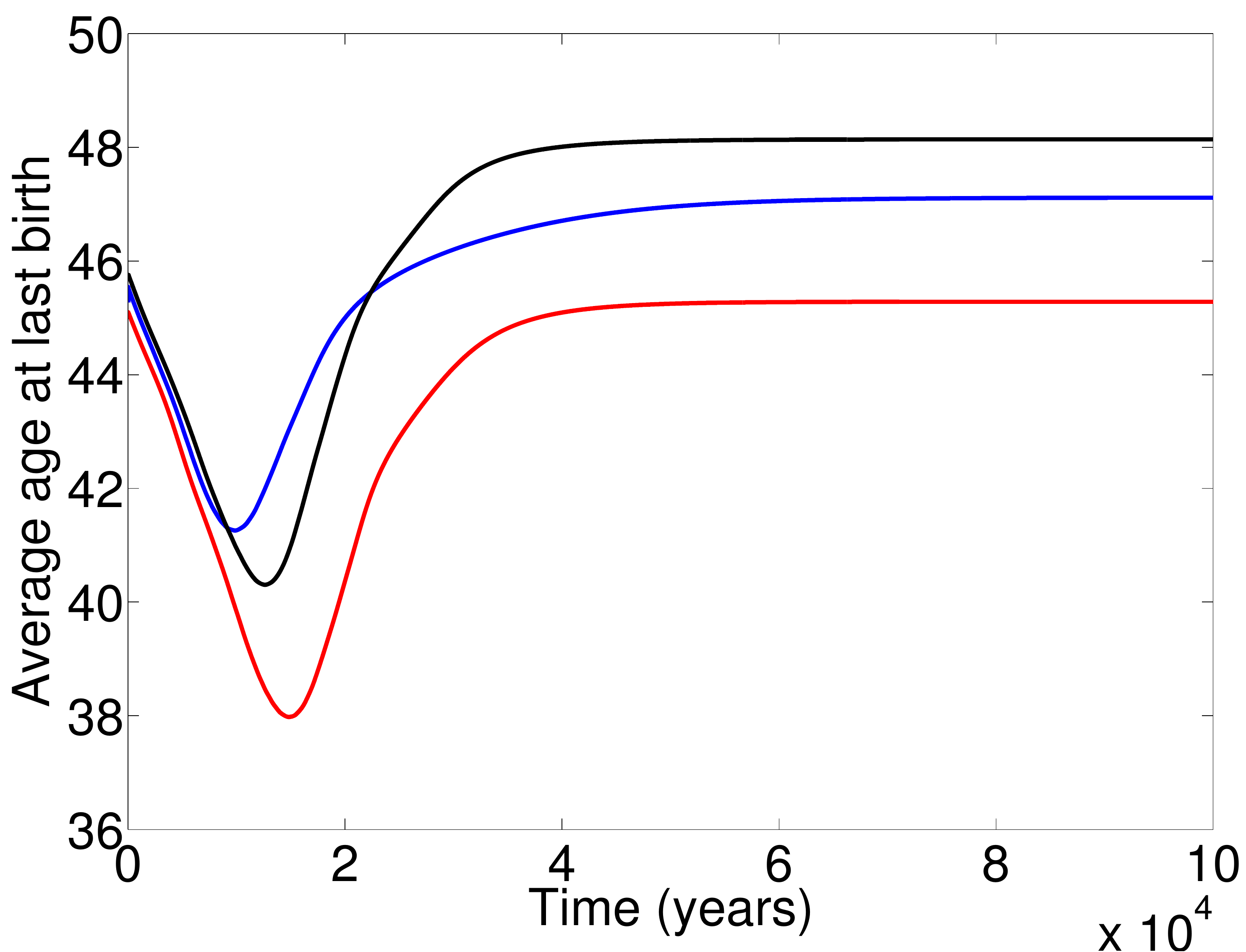}}
  \hspace{5mm}
  \subfloat[]{\label{sensitivity_tau1_e0}\includegraphics[width=0.45\textwidth]{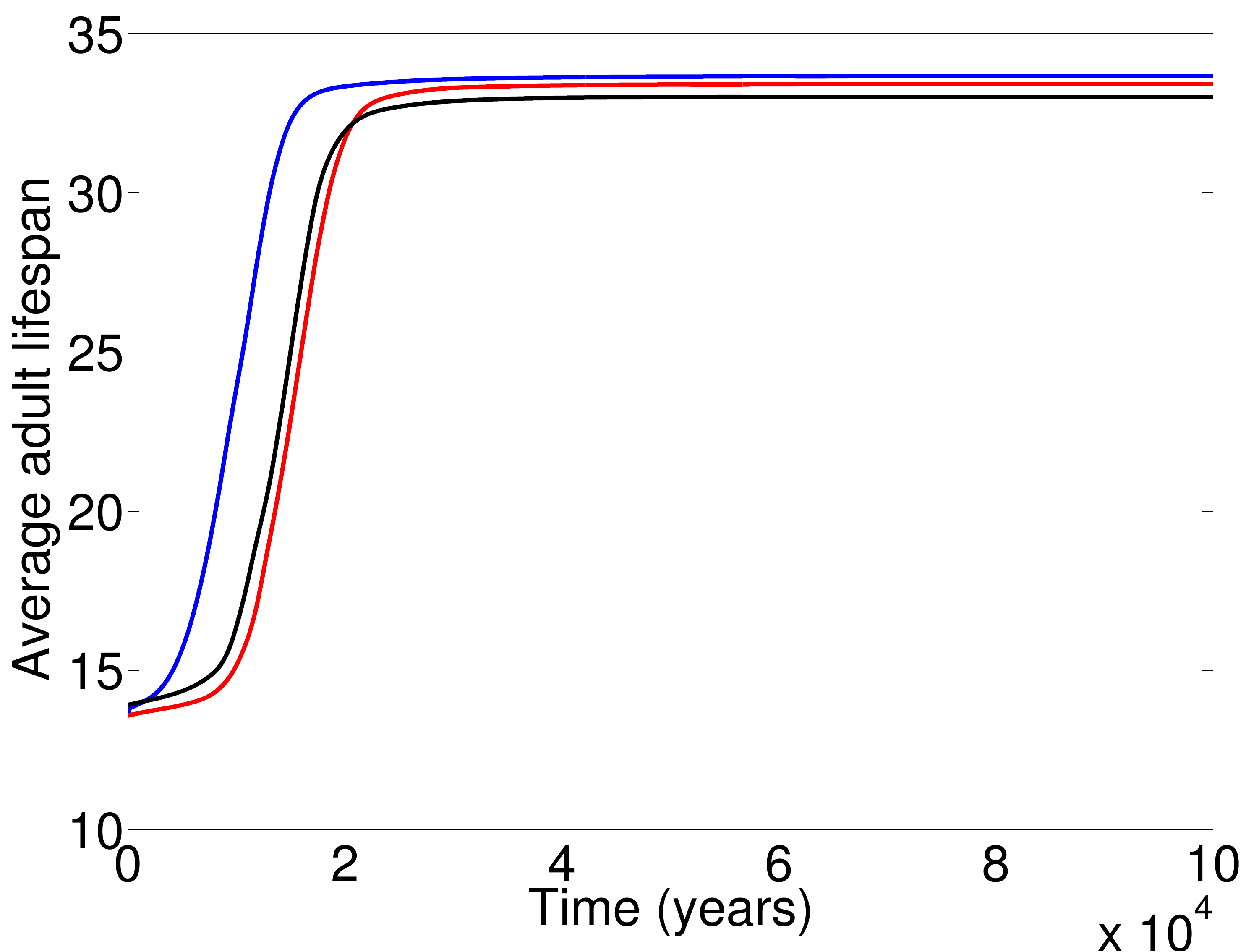}}}
  \caption{Time evolutions of the system with grandmothering with $0.8\tau_1(L)$ (red), $0.9\tau_1(L)$ (blue) and $\tau_1(L)$ (black). The initial condition used was the corresponding equilibria of the system without grandmothering, shown in Figure \ref{sensitivity_tau1_chimps}.}
  \label{sensitivity_tau1_GM}
\end{figure}

\begin{figure}[H]
  \centerline{
  \includegraphics[width=0.45\textwidth]{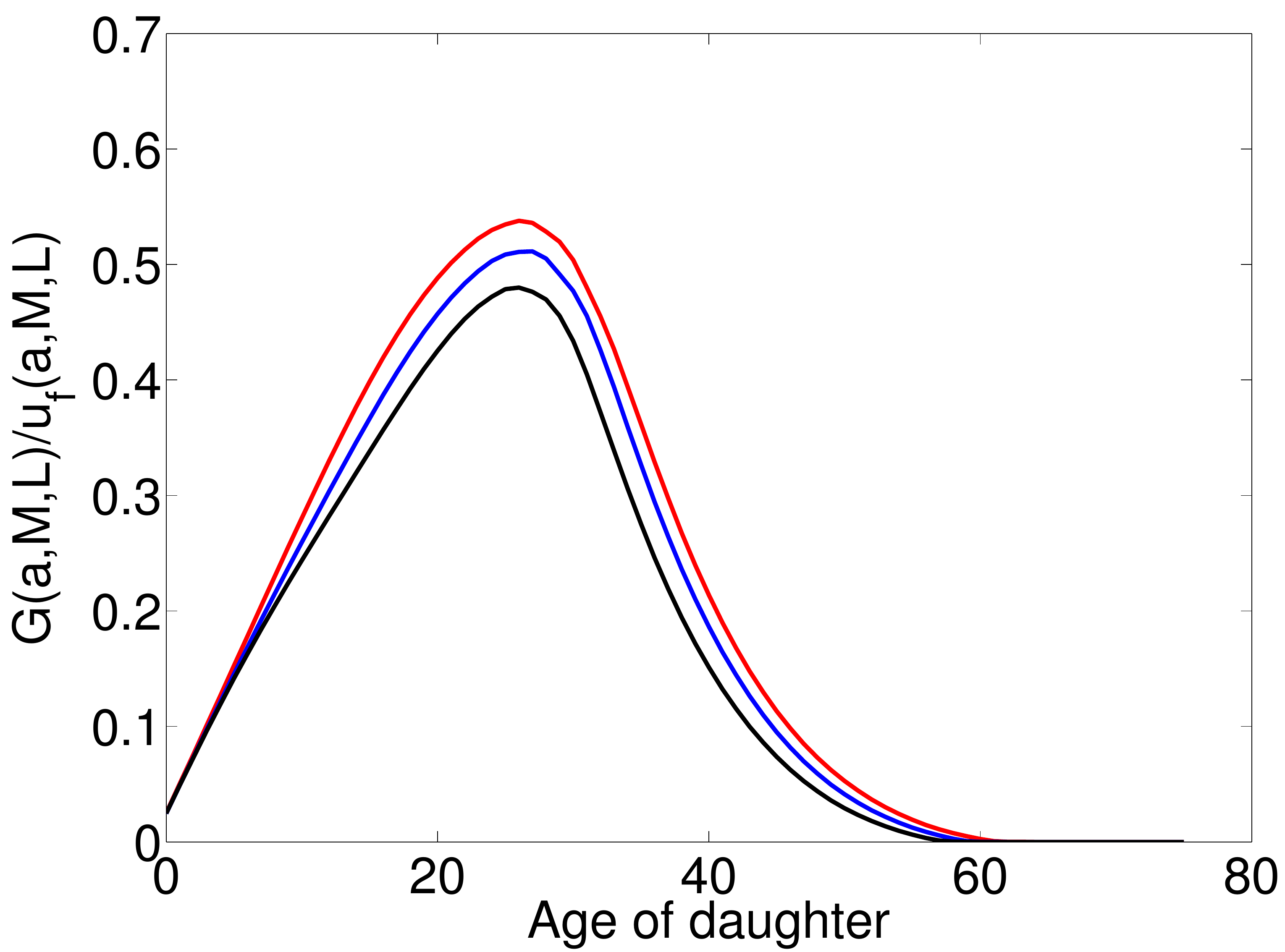}}
  \caption{The proportion of females of any age with a live mother with age $a \in [\tau_2(M),\tau_3]$ (that is, $G(a,M,L)/u_f(a,M,L)$) at the hunter-gatherer equilibrium with with $0.8\tau_1(L)$ (red), $0.9\tau_1(L)$ (blue) and $\tau_1(L)$ (black).}
  \label{sensitivity_tau1_daughterscovered}
\end{figure}

A comparison of the great ape-like equilibrium and the hunter-gatherer equilibrium with ages of female sexual maturity $0.8\tau_1(L)$, $0.9\tau_1(L)$ and $\tau_1(L)$ are shown in Figures \ref{sensitivity_tau1_chimps}-\ref{sensitivity_tau1_GM}. For both with and without grandmothering, a delayed female sexual maturity increases both average adult lifespan and age at last birth, although for the case without grandmothering the increases are small (see Figure \ref{sensitivity_tau1_chimps}). For $1.1\tau_1(L)$ (not shown), the average adult lifespan increases past a critical value where the population is able to escape the basin of attraction of the great ape-like equilibrium. This leads to the population increasing its average adult lifespan to approximately 35, at which point it goes extinct. These findings for average adult lifespan under variation of the age of female sexual maturity generally agree with the results of Kim et al. \cite{116}, who obtain similar results using an agent-based model. A decrease in $\tau_1(L)$ results in a particularly large increase in average age at last birth for the case with grandmothering because the introduction of younger sexually mature females leads to post-fertile females supporting more daughters, since post-fertile females support all their daughters provided they are sexually mature. This results in greater grandmother effects in the system which in turn generates greater selection for a lower age at last birth to further increase the number of grandmothers. This is shown in Figure \ref{sensitivity_tau1_daughterscovered}.

\subsection{Late mortality rate}
\label{Senescent_mortality}

\begin{figure}[H]
  \hspace{1mm}\centerline{
  \subfloat[]{\label{sensitivity_normalmus_chimps_M}\includegraphics[width=0.45\textwidth]{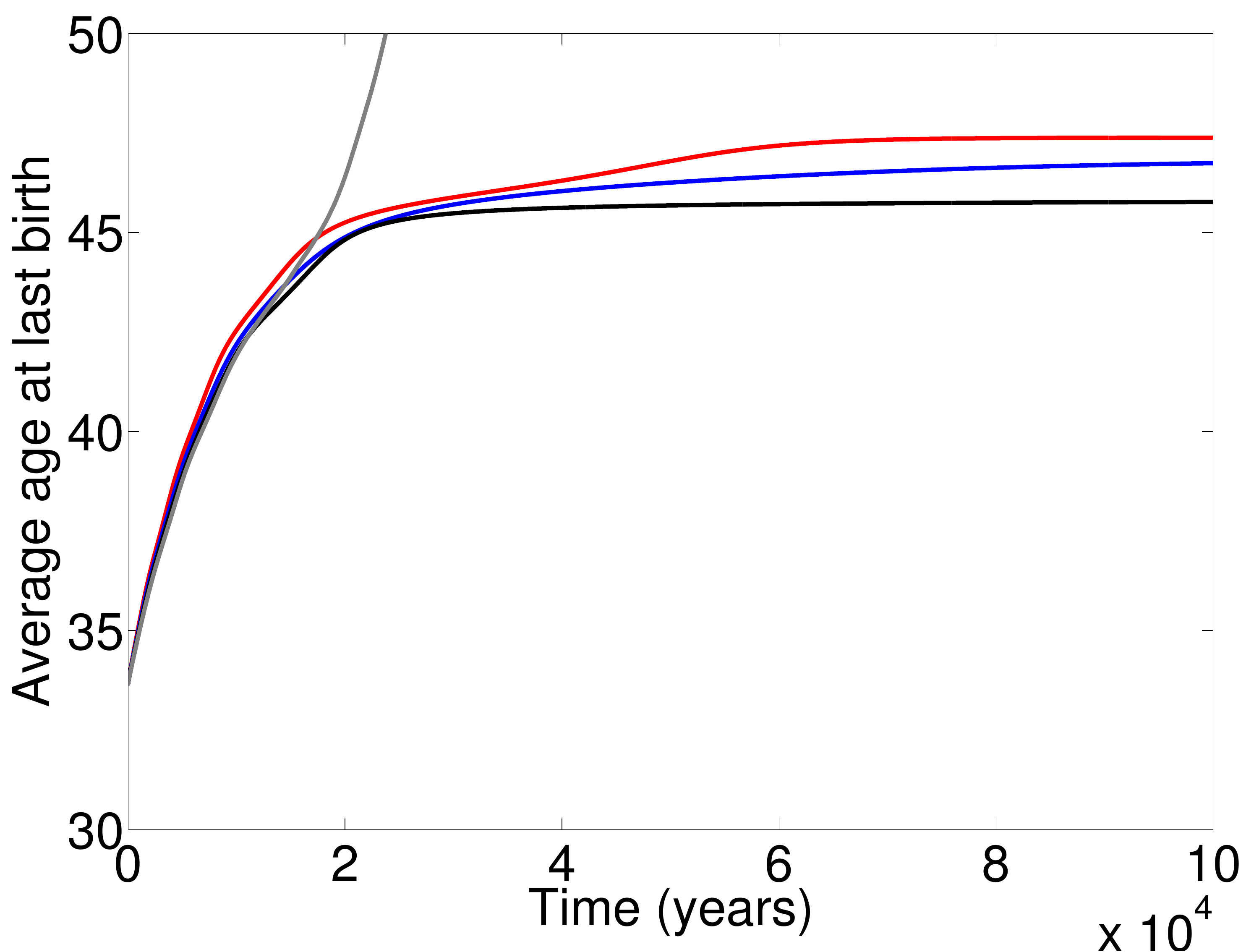}}
  \hspace{5mm}
  \subfloat[]{\label{sensitivity_normalmus_chimps_L}\includegraphics[width=0.45\textwidth]{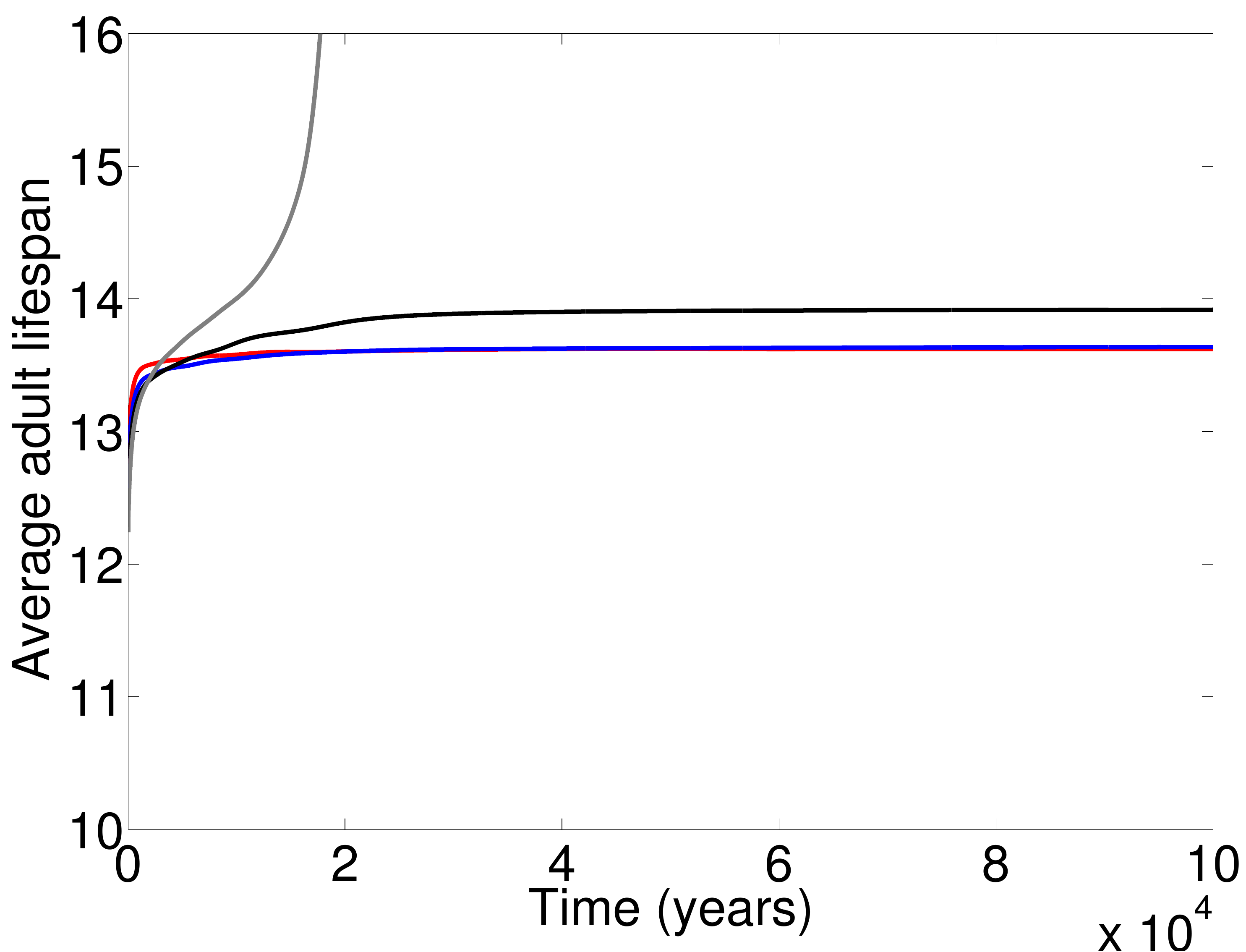}}}
  \caption{Time evolutions of the system without grandmothering with $0.9\mu_s$ (red), $0.95\mu_s$ (blue) and $\mu_s$ (black), $1.05 \mu_s$ (grey). The simulation with $1.05\mu_s$ reaches an equilibrium with average age at last birth of 67.25 and average adult lifespan of 33.5, however the population goes extinct as the net reproductive rate is negative.}
  \label{sensitivity_normal_mus_chimps}
\end{figure}

\begin{figure}[H]
  \hspace{1mm}\centerline{
  \subfloat[]{\label{sensitivity_normalmus_GMs_M}\includegraphics[width=0.45\textwidth]{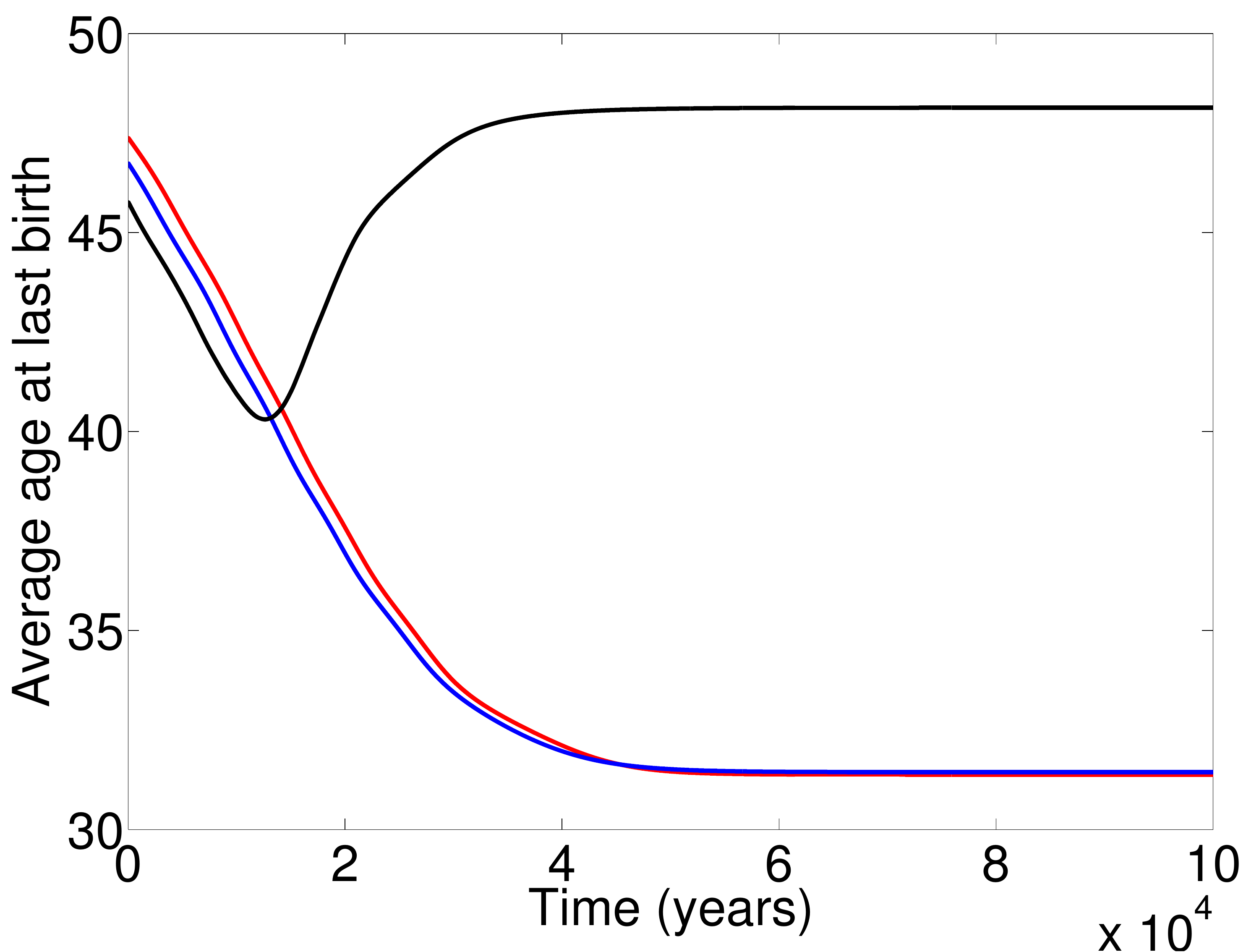}}
  \hspace{5mm}
  \subfloat[]{\label{sensitivity_normalmus_GMs_L}\includegraphics[width=0.45\textwidth]{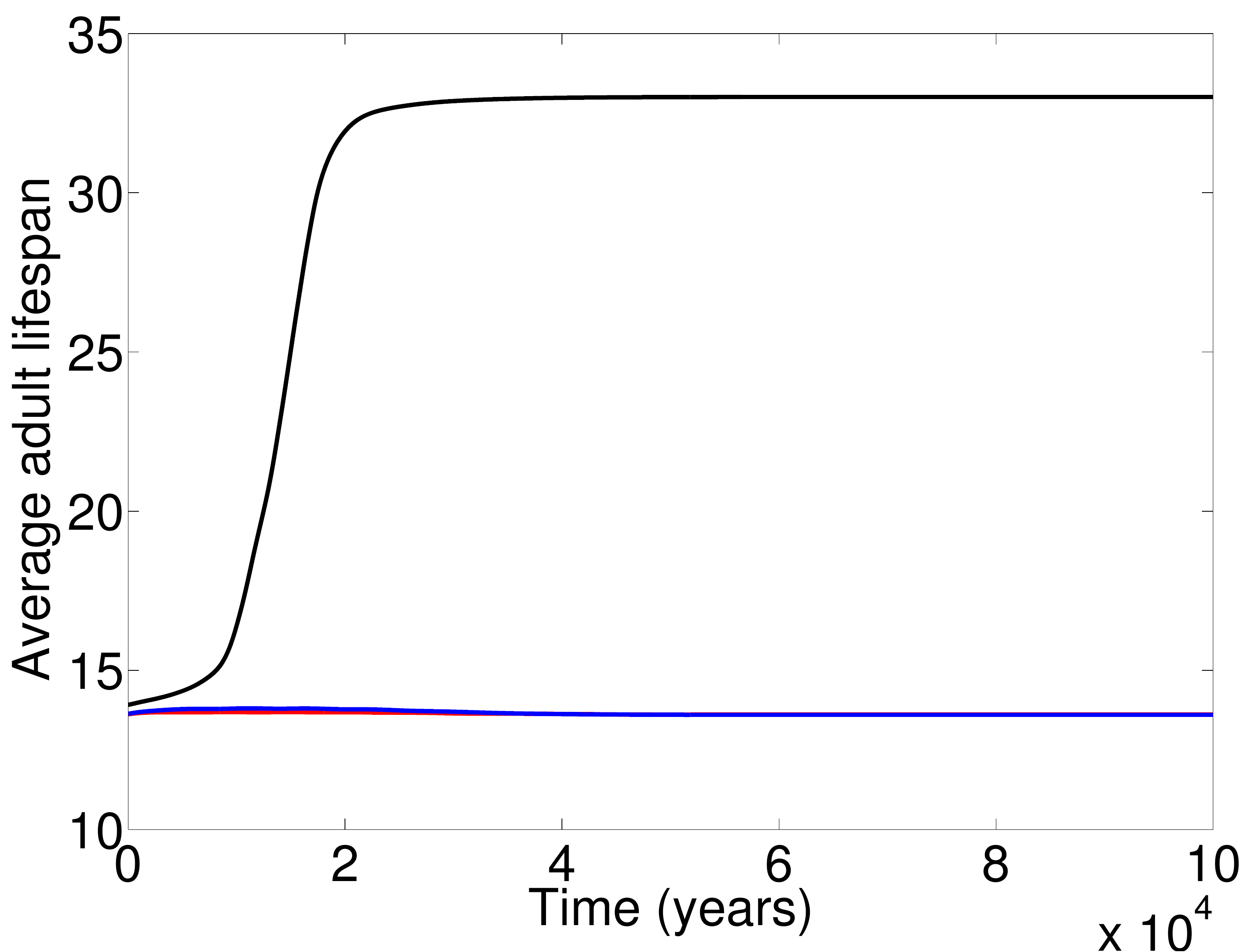}}}
  \caption{Time evolutions of the system with grandmothering with $0.9\mu_s$ (red), $0.95\mu_s$ (blue) and $\mu_s$ (black). The initial condition used was the corresponding equilibria of the system without grandmothering, shown in Figure \ref{sensitivity_mus_chimps}. We do not include the $1.05 \mu_s$ case here as the population goes extinct without grandmothering and even with grandmothering the net reproductive rate is still negative.}
  \label{sensitivity_normal_mus_GM}
\end{figure}

\begin{figure}[H]
  \hspace{1mm}\centerline{
  \subfloat[]{\label{sensitivity_mus_chimps_M}\includegraphics[width=0.45\textwidth]{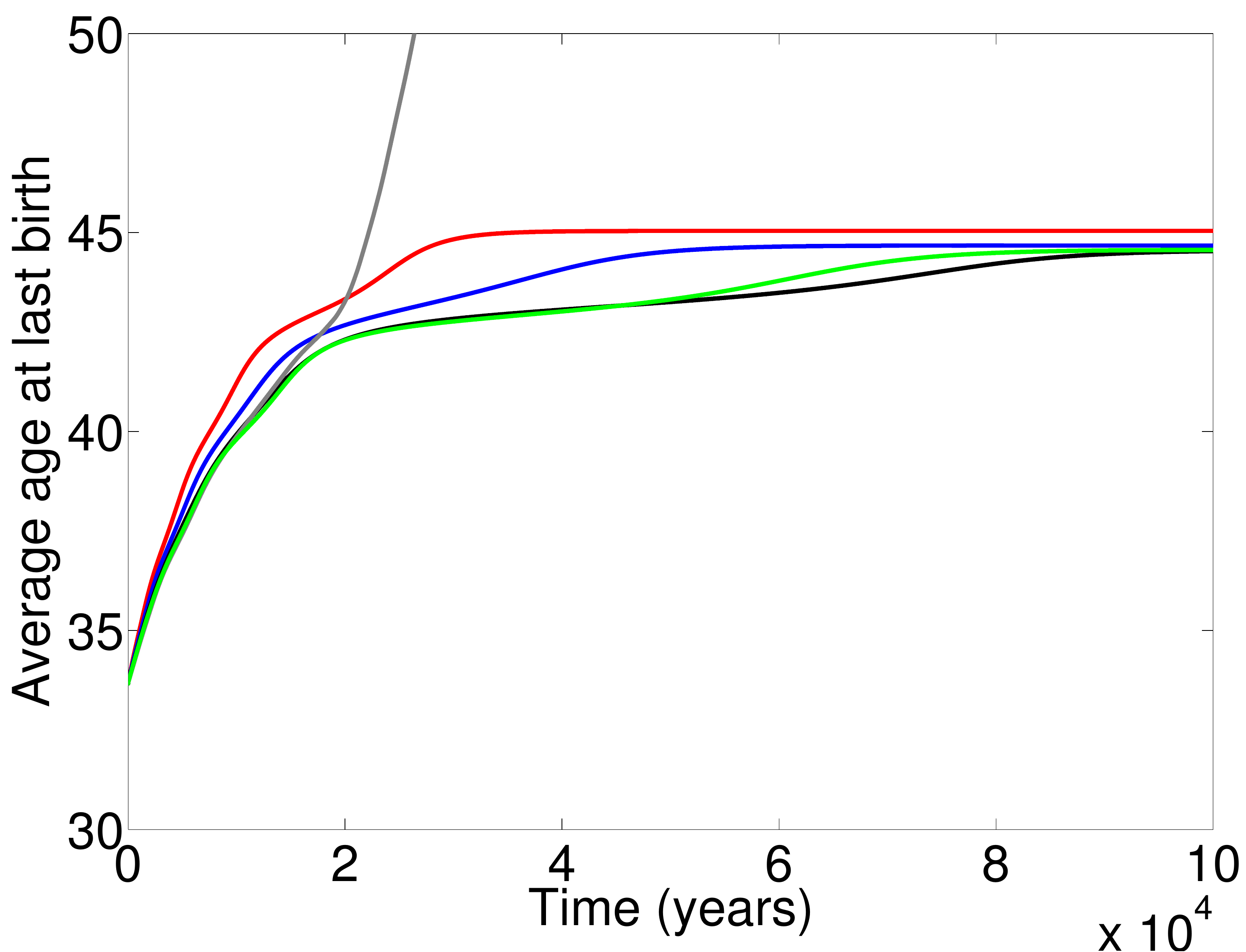}}
  \hspace{5mm}
  \subfloat[]{\label{sensitivity_mus_chimps_L}\includegraphics[width=0.45\textwidth]{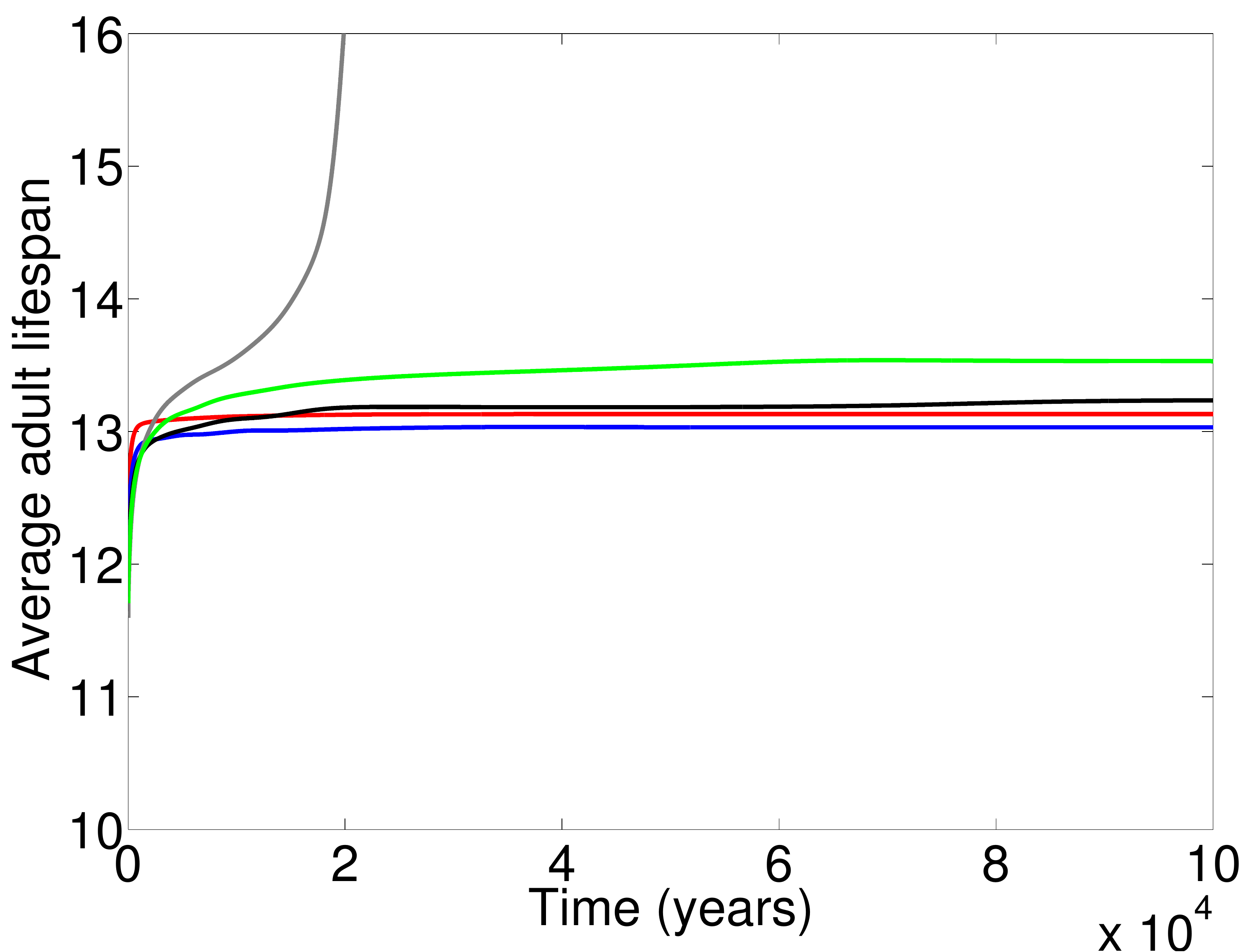}}}
  \caption{Time evolutions of the system without grandmothering with $\mu_s$ (red), $1.1\mu_s$ (blue) and $1.2\mu_s$ (black), $1.25 \mu_s$ (green) and $1.3 \mu_s$ (grey). The male ferility-longevity tradeoff function used is $\phi_3(L)$. The simulation with $1.3\mu_s$ reaches an equilibrium with average age at last birth of 67.25 and average adult lifespan of 33.5, however the population goes extinct as the net reproductive rate is negative.}
  \label{sensitivity_mus_chimps}
\end{figure}

\begin{figure}[H]
  \hspace{1mm}\centerline{
  \subfloat[]{\label{sensitivity_mus_GMs_M}\includegraphics[width=0.45\textwidth]{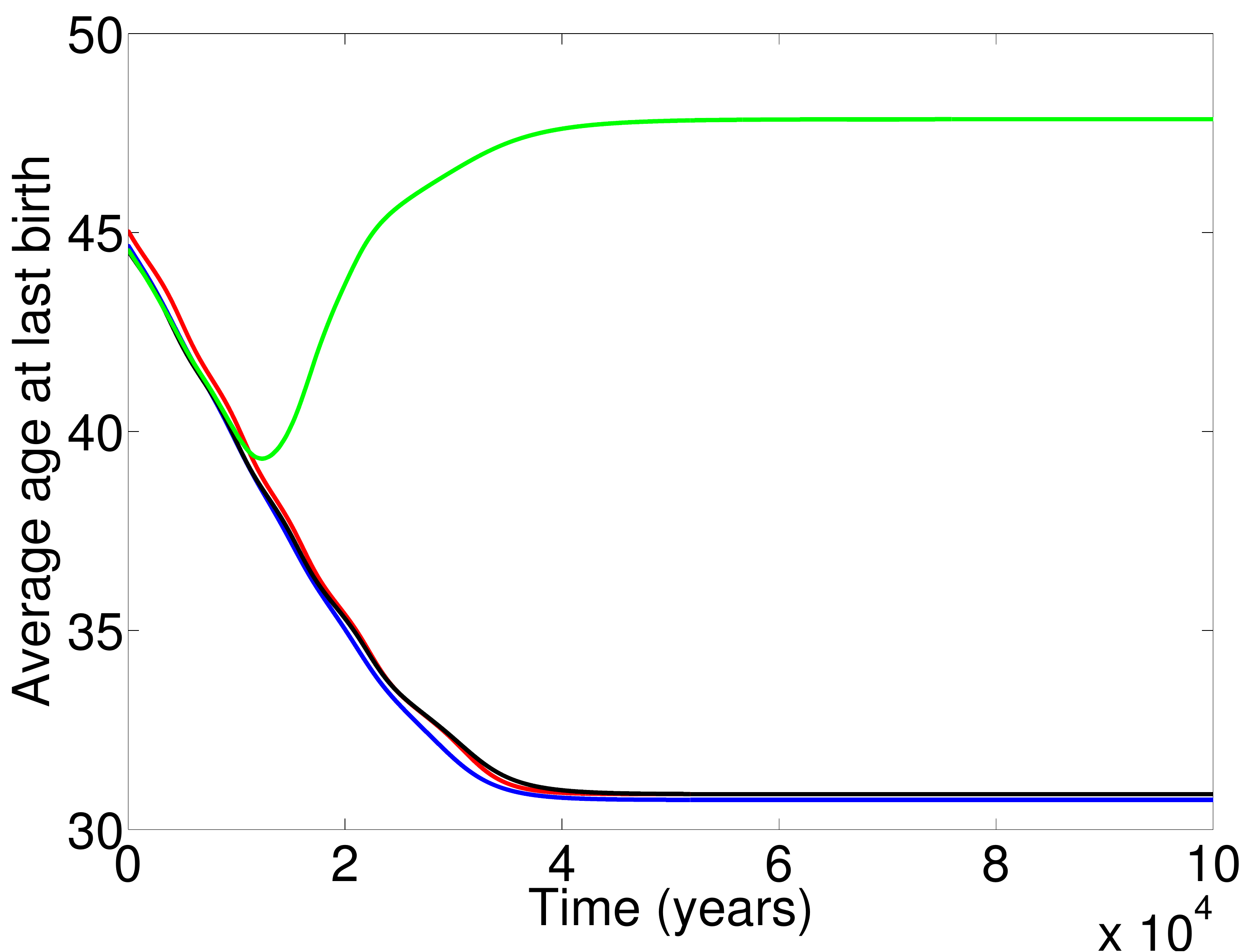}}
  \hspace{5mm}
  \subfloat[]{\label{sensitivity_mus_GMs_L}\includegraphics[width=0.45\textwidth]{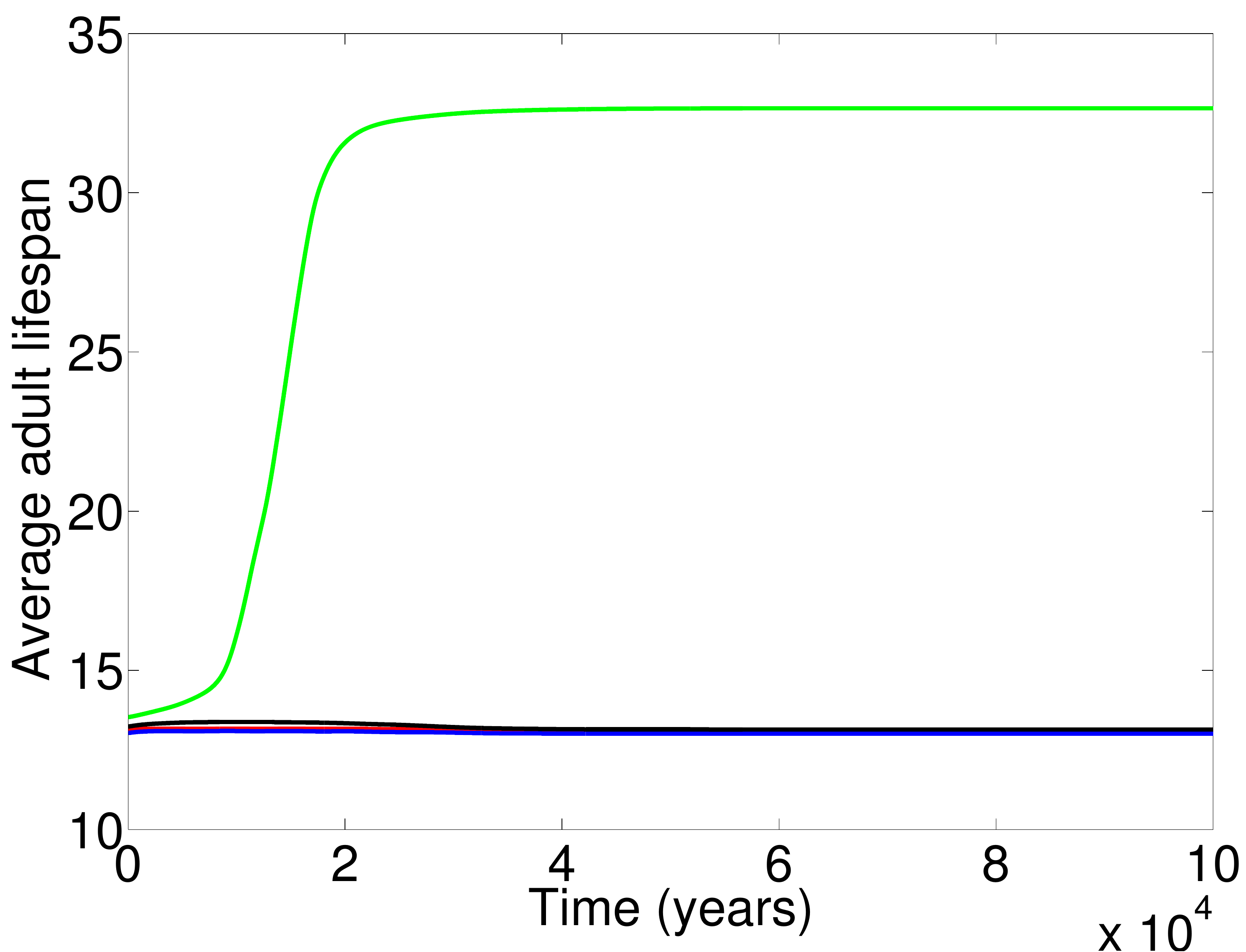}}}
  \caption{Time evolutions of the system with grandmothering with with $\mu_s$ (red), $1.1\mu_s$ (blue) and $1.2\mu_s$ (black) and $1.25 \mu_s$ (green). The male ferility-longevity tradeoff function used is $\phi_3(L)$. The initial condition used was the corresponding equilibria of the system without grandmothering, shown in Figure \ref{sensitivity_mus_chimps}. We do not include the $1.3 \mu_s$ case here as the population goes extinct without grandmothering and even with grandmothering the net reproductive rate is still negative.}
  \label{sensitivity_mus_GM}
\end{figure}

In Figures \ref{sensitivity_normal_mus_chimps} and \ref{sensitivity_normal_mus_GM} we examine the effect of the base rate of late mortality $\mu_s$ by comparing the equilibria generated when we use $0.9\mu_s$, $0.95\mu_s$, $\mu_s$ and $1.05\mu_s$. Without grandmothering, an increase in the late mortality rate results in an increase in average adult lifespan; this is due to there being fewer post-fertile females (who add nothing to the growth of the population in the case without grandmothering), which makes room for more fertile females as longevity increases. From Figure \ref{sensitivity_normal_mus_chimps}, we observe that only the solutions generated by $0.9\mu_s$, $0.95\mu_s$ and $\mu_s$ have a basin of attraction at the great ape-like equilibrium. For $1.05 \mu_s$, the population can escape the basin of attraction and progress towards the hunter-gatherer equilibrium without grandmothering, leading to a negative net reproductive rate and hence going extinct. Figure \ref{sensitivity_mus_chimps}, which uses $\phi_3(L)$ as the male fertility-longevity tradeoff function, shows similar results; the solutions generated by $\mu_s$, $1.1\mu_s$, $1.2\mu_s$, $1.25\mu_s$ are trapped in the basin of attraction at the great ape-like equilibrium, whereas the solution generated by $1.3\mu_s$ can escape it, but goes extinct.
\\
\\
The transition between great ape-like longevities and hunter-gatherer longevities, without extending the end of fertility, is not automatic; from Figure \ref{sensitivity_normal_mus_GM} and Figure \ref{sensitivity_mus_GM} we see that this transition is only possible for $\mu_s$ and $1.25\mu_s$ respectively. This demonstrates the sensitivity and non-linearity of this transition between the two equilibria; even though a lower base late mortality rate increases the number of grandmothers, a higher base late mortality rate is needed to ensure that the population can escape the basin of attraction when grandmothering is introduced. These results imply that the parameter space in which such a transition is possible is small, which agrees with Kim et al. \cite{116} who suggest that ``great apes exist at the brink of a critical window" where a transition to the hunter-gatherer equilibrium is possible with grandmothering. For male fertility-longevity tradeoff functions $\phi(L)$ and $\phi_3(L)$, we measure this window to be within the interval $L=[0.1415,0.1548]$ and $[0.1512,0.1635]$ respectively. Thus, changes in the shape of the male fertility-longevity tradeoff function can affect the position of the critical window; this is due to the need for the great ape-like equilibrium to be close to the plateau in the male fertility-longevity tradeoff function (see Figure \ref{sensitivity_matingfail}). 
\\
\\
The average age at last birth is unrealistically low when the population cannot make the transition to the hunter-gatherer equilibrium, as shown in Figure \ref{sensitivity_phi_GM}, Figure \ref{sensitivity_normal_mus_GM} and Figure \ref{sensitivity_mus_GM}, which suggests grandmothering is always beneficial to the net reproductive rate of the population, regardless of whether it can make the transition. This is due to the assumption of a fixed $m$ (benefit from grandmothers) across $M$ and $L$; $m$ in reality should increase with $L$, since the period in which an individual is weaned, but still dependent, increases with $L$ according to Charnovian assumptions. The selection for a post-fertile lifespan stems from the assumption that grandmothers provision weaned grandchildren who are not independent (i.e. they cannot effectively forage for themselves); without the grandmother, these grandchildren would be dependent on their mothers which would increase the average interbirth interval as the mother cannot have next offspring without risking the survival of the current dependent offspring. Thus, in environments where just weaned offspring are able to effectively forage for themselves, selection for a post-fertile lifespan is expected to be weak to the point where the population growth rate is higher when females are fertile their entire life. This is supported by Figure \ref{sensitivity_B_menopause}, which shows that a lower $m$ can greatly increase the average age at last birth.

\section{Discussion}

In this study, we have used a two-sex PDE model to find that the fitness benefits gained from grandmothers can drive the evolution of longevity without extending the end of fertility. Other quantitative studies that have addressed this problem include Kim et al. \cite{116}, Kim et al. \cite{117}, Pavard \& Branger \cite{118} and Kachel \cite{115}. The key features of our study are (1) it is a two-sex model, (2) we allow for individuals to mutate (and evolve) in age at last birth, in addition to longevity. We simplify the complex physiology of terminal fertility (well characterised in women, see, eg. Velde \& Pearson \cite{123}), using $M$ as the label for average age at last birth to suggest the familiar label menopause. Additional assumptions include (3) grandmothers only support their daughters and (4) the population is expressed in terms of population density. We use findings from Kim et al. \cite{116}, Kim \cite{117} and Field et al. \cite{168} that identify two locally stable equilibria, corresponding to great ape-like lifespans and hunter gatherer-like lifespans, under the effects of grandmothering and investigate selection on both longevity and the end of female fertility around those equilibria. To our knowledge, the only other study, apart from ours, which has attempted to show the evolutionary trajectories of both longevity and age at last birth together in a two-sex model, is by Kachel et al. \cite{115}, however their results were inconclusive as they had not assumed a male fertility-longevity tradeoff, which resulted in overriding selection for increased longevity through males (see Kachel et al. \cite{113} and Hawkes et al. \cite{114}).
\\
\\
{Mammalian life-histories lie on a fast-slow continuum, in which life-history traits such as lifespan and age at maturity are positively correlated with each other, but negatively correlated with fertility. In our model, we have let the single variable $L$ parametrise these life-history traits, which follows from Charnov's theory on mammalian life-history invariants \cite{129}. As such, $L \in [0,1]$ acts as a measure of the fast-slow continuum. The interconnected nature of these life-history traits is based on the principle that energy invested towards reproduction results in less energy spent on somatic maintenance -- The fast end of the continuum is characterised by high fertility and low age of maturation, but high mortality rates, whereas the slow end is the opposite; it is characterised by reduced fertility and increased age at maturation, but lower mortality rates \cite{129}. Grandmothering allows individuals to invest more into somatic maintenance; the subsidies from these post-reproductive years in turn allow fertile mothers to invest less in their offspring even though these offspring are dependent longer. Thus, grandmothering effects allow for both perks of high fertility and low mortality rates to be achieved, where the cost is that a grandmother forgoes the opportunity to have her next offspring. 
\\
\\
Predictably, whether this cost of grandmothering is prohibitively expensive from an evolutionary standpoint depends heavily on the male and female tradeoff functions amongst other parameter values as shown in the Section \ref{sensitivitysect}. We have based the parameter values on empirical data so as to ensure the two equilibria have accurate life-history properties, such as mean age of last birth, mean life expectancy, mean age at first birth, post-reproductive representation and age profile. However, as there is a lack of data for the intermediate species during human evolution, the form of the male and female fertility-longevity tradeoff functions for $L$ in-between 0.2 and 0.8 are essentially unknown. Although these tradeoff functions are free parameters, we find that the shapes of these tradeoff functions must be somewhat precise; as we have shown in Section \ref{sensitivitysect}, small deviations in the male fertility-longevity tradeoff function alone can induce a bifurcation in the system whereby there is no evolutionary trajectory between the two equilibria. We refer the reader to Section \ref{sensitivity_male} for an explanation of the specific shape required for the male fertility-longevity tradeoff. We found that small deviations in the female fertility-longevity tradeoff do not affect the existence of the evolutionary trajectory between equilibria; however they have a large influence on the life-history trait values at both equilibria. Thus, the conditions on the shape of the female fertility-longevity tradeoff can be viewed as less stringent than the corresponding male tradeoff.}
\\
\\
A pivotal distinction of our model is that we do not fix the end of female fertility; this assumption has been made in previous studies due to the fact that human females and other great apes share a similar average age at last birth. However, since the key difference is that human females routinely survive beyond their fertile years, such an assumption does not answer the question of whether menopause is an adaptation, resulting from the grandmother hypothesis, or merely an epiphenomenon due to either a physiological trade-off favouring efficient early reproduction or simply a result of adult lifespans increasing beyond the female's supply of eggs or ability to sustain ovulatory cycles (see Peccei \cite{147} for a discussion). Although the adaptation hypothesis and the physiological tradeoff hypothesis are not necessarily mutually exclusive, the main difference is that in the latter, menopause is by-product of declining fertility rates before the age at last birth. Peccei \cite{147} notes that ``whether menopause is the result of selection for efficient early reproductive or selection for a postreproductive lifespan is exceedingly difficult to tease apart". It is possible that it is a mix of both, but we show that selection for a post-fertile lifespan, whereby mothers increase the fertility rate of their daughters by provisioning for their grandchildren, is alone enough to maintain an average age at last birth of 45-50 years.
\\
\\
By assuming that mothers only support their daughters, the model is required to, in a sense, track lineages. This adds considerable complexity to the model, which in turn makes the simulations computationally more demanding. For example, on a single core of a CPU (unfortunately, we could not find an efficient method to implement CUDA to solve our system due to the non-local nature of the mating mechanism in the model) our model took 3 hours to simulate $100,000$ years without grandmothering, but with grandmothering (i.e. the model is required to ``track lineages") the run time increased to 24 hours for the same number of years. A run time of 24 hours, although by no means considered long compared to other computationally intensive problems, is unwieldy for both calibration of the model and sensitivity analyses. This is perhaps the disadvantage of using a PDE approach, as opposed to an agent-based approach to track the evolution of traits in a population. This was noted by Kim et al. \cite{116}, who found that the run time for an agent-based model was much shorter than a PDE model. However, a clear advantage of deterministic models is being able to detect small quantitative changes in the system, resulting from small parameter changes, without the need to resort to Monte Carlo methods as required with agent-based models.
\\
\\
The time transition between the great ape-like equilibrium and the hunter gatherer-like equilibrium in our model takes approximately 30,000 years, assuming that offspring inherit the mean of their parents' $M$ and $L$ values with a probability of $5\%$ to mutate a maximum of $\pm 0.05$ in either $M$ or $L$. One possible reason for this unrealistic transition speed is the deterministic formulation of the model. Kim et al.'s \cite{117} deterministic model also yielded quick transition speeds (20,000 to 50,000 years) between equilibria with the same probability for a mutational shift in longevity. A comparison of the deterministic model by Kim et al. \cite{117} with the probabilistic model by Kim et al. \cite{116} shows that including stochasticity slowed transition times by almost an order of magnitude. Another important difference is that Kim et al. \cite{116} and Kim et al. \cite{117} both assumed indiscriminate grandmothering, in contrast to our study which assumes that grandmothers can only support their daughters, which potentially affects the speed of the transition.
\\
\\
While indiscriminate grandmothering is a useful assumption to bypass the complexity introduced by lineage tracking, it results in many more eligible grandmothers (as even post-fertile females whose daughters have died can support fertile females) than a model where mothers only support their own daughters, and so it is unclear whether the former overestimates grandmother effects and if so, by how much. Moreover, indiscriminate grandmothering is a form of altruism, as females who only support their own daughters have more grandchildren on average, and so it is questionable whether this is an evolutionarily stable strategy \cite{168}. Explanations for the evolution of altruism, and the necessary assumptions, have been widely discusses in the literature. Hamilton \cite{159,160} first suggested that viscosity in populations (i.e. individuals have limited dispersal) promotes altruistic behaviour as interactions between individuals tend to be with kin. However, various modelling studies have suggested that this is not necessarily true because as the genetic relatedness of neighbours increases, so too does the competition between kin (these are reviewed in \cite{161}). Whether this competition between kin completely inhibits or merely reduces the selection for altruism has been shown to be heavily dependent on the assumptions of the model. For example, Wilson et al. \cite{166} used an agent-based model, where females interact with neighbouring females on a grid and their offspring disperse locally, showed that local population regulation negated the benefits from altruism; Taylor \cite{163,164} used an inclusive fitness approach to show the same result as Wilson et al. \cite{166}; Gardner \& West \cite{162} find that budding (groups of individuals dispersing together) enables selection for altruism as it reduces local competition and increases the degree of kinship; Mittledorf \& Wilson \cite{165} find that allowing for elastic population densities, where patches of altruists are supported at higher densities than patches of non-altruists, can support the selection for altruism.
\\
\\
Another approach to modelling grandmothering effects in a population is to remove males from the model, to create a single-sex model, which simplifies the model and hence resulting dynamics and analysis (for examples, see \cite{128,118,168}). These advantages may not outweigh the cost of ignoring sexual selection and sexual conflict (see Arnqvist \& Rowe \cite{132} and Kokko \& Jennions \cite{131}) which could play an important role in our lineage, especially since male fertility continues to older ages in humans, skewing the sex ratio in the fertile ages towards the males \cite{169}. Single-sex models can suggest that selection for higher longevities maximises the net reproductive rate of the population, but our study and also Kim et al. \cite{116} show that this sexual conflict can promote selection for longevity values that are detrimental to the net reproductive rate of the whole population and that whether a population can evolve from great ape-like longevities to hunter-gatherer longevities is sensitive to the shape of the male fertility tradeoff function.

\begin{acknowledgements}
The work of MHC was supported by the Australian Postgraduate Award. PSK was supported by the Australian Research Council, Discovery Early Career Research Award (DE120101113), and the University of Sydney, Bridging Support Grant (BSG176616). 
\end{acknowledgements}
\bibliographystyle{apalike}
\bibliography{references}

\begin{thebibliography}{}

\bibitem[Alberts et~al., 2013]{127}
Alberts, S.~C., Altmann, J., Brockman, D.~K., Cords, M., Fedigan, L.~M., Pusey,
  A., Stoinski, T.~S., Strier, K.~B., Morris, W.~F., and Bronikowski, A.~M.
  (2013).
\newblock {{R}eproductive aging patterns in primates reveal that humans are
  distinct}.
\newblock {\em Proc. Natl. Acad. Sci. U.S.A.}, 110(33):13440--13445.

\bibitem[Arnqvist and Rowe, 2005]{132}
Arnqvist, G. and Rowe, L. (2005).
\newblock {\em Sexual Conflict}.
\newblock Monographs in behavior and ecology. Princeton University Press.

\bibitem[Blurton~Jones et~al., 2002]{143}
Blurton~Jones, N.~G., Hawkes, K., and O'Connell, J.~F. (2002).
\newblock {{A}ntiquity of post-reproductive life: are there modern impacts on
  hunter-gatherer post-reproductive life spans?}
\newblock {\em American Journal of Human Biology}, 14(2):184--205.

\bibitem[Boesch et~al., 2006]{146}
Boesch, C., Kohou, G., Nene, H., and Vigilant, L. (2006).
\newblock {{M}ale competition and paternity in wild chimpanzees of the
  {T}a{\"i} forest}.
\newblock {\em American Journal of Physical Anthropology}, 130(1):103--115.

\bibitem[Charnov, 1993]{129}
Charnov, E.~L. (1993).
\newblock {\em Life History Invariants: Some Explorations of Symmetry in
  Evolutionary Ecology}.
\newblock Life History Invariants: Some Explorations of Symmetry in
  Evolutionary Ecology. Oxford University Press.

\bibitem[Coale and Demeny, 1968]{153}
Coale, A. and Demeny, P. (1968).
\newblock {\em Regional Model Life Tables and Stable Population}.
\newblock Princeton University Press, New Jersey.

\bibitem[Coxworth et~al., 2015]{169}
Coxworth, J.~E., Kim, P.~S., McQueen, J.~S., and Hawkes, K. (2015).
\newblock Grandmothering life histories and human pair bonding.
\newblock {\em Proceedings of the National Academy of Sciences},
  112(38):11806--11811.

\bibitem[Croft et~al., 2015]{154}
Croft, D.~P., Brent, L.~J., Franks, D.~W., and Cant, M.~A. (2015).
\newblock The evolution of prolonged life after reproduction.
\newblock {\em Trends in Ecology and Evolution}, 30(7):407--416.

\bibitem[Field et~al., 2015]{168}
Field, J.~M., Hawkes, K., and Kim, P.~S. (2015).
\newblock A continuous model of grandmothering: thresholds, evolutionary
  stability, and who gets help.
\newblock {\em In review}.

\bibitem[Gardner and West, 2006]{162}
Gardner, A. and West, S.~A. (2006).
\newblock {{D}emography, altruism, and the benefits of budding}.
\newblock {\em Journal of Evolutionary Biology}, 19(5):1707--1716.

\bibitem[Gurven and Kaplan, 2007]{130}
Gurven, M. and Kaplan, H. (2007).
\newblock Longevity among hunter-gatherers: A cross-cultural examination.
\newblock {\em Population and Development Review}, 33(2):321--365.

\bibitem[Hamilton, 1964a]{159}
Hamilton, W.~D. (1964a).
\newblock {{T}he genetical evolution of social behaviour. {I}}.
\newblock {\em Journal of Theoretical Biology}, 7(1):1--16.

\bibitem[Hamilton, 1964b]{160}
Hamilton, W.~D. (1964b).
\newblock {{T}he genetical evolution of social behaviour. {I}{I}}.
\newblock {\em Journal of Theoretical Biology}, 7(1):17--52.

\bibitem[Hawkes, 2003]{120}
Hawkes, K. (2003).
\newblock Grandmothers and the evolution of human longevity.
\newblock {\em American Journal of Human Biology}, 15(3):380--400.

\bibitem[Hawkes, 2004]{133}
Hawkes, K. (2004).
\newblock {{H}uman longevity: the grandmother effect}.
\newblock {\em Nature}, 428(6979):128--129.

\bibitem[Hawkes et~al., 2011]{114}
Hawkes, K., Kim, P.~S., Kennedy, B., Bohlender, R., and Hawks, J. (2011).
\newblock A reappraisal of grandmothering and natural selection.
\newblock {\em Proceedings of the Royal Society of London B: Biological
  Sciences}, 278(1714):1936--1938.

\bibitem[Hawkes et~al., 1995]{155}
Hawkes, K., O'Connell, J.~F., and Blurton~Jones, N.~G. (1995).
\newblock Hadza children's foraging: Juvenile dependency, social arrangements,
  and mobility among hunter-gatherers.
\newblock {\em Current Anthropology}, 36(4):688--700.

\bibitem[Hawkes et~al., 1997]{156}
Hawkes, K., O'Connell, J.~F., and Blurton~Jones, N.~G. (1997).
\newblock Hadza women's time allocation, offspring provisioning, and the
  evolution of long postmenopausal life spans.
\newblock {\em Current Anthropology}, 38(4):551--577.

\bibitem[Hawkes et~al., 1998]{157}
Hawkes, K., O'Connell, J.~F., Blurton~Jones, N.~G., Alvarez, H., and Charnov,
  E.~L. (1998).
\newblock Grandmothering, menopause, and the evolution of human life
   histories.
\newblock {\em Proceedings of the National Academy of Sciences},
  95(3):1336--1339.

\bibitem[Hawkes et~al., 2009]{124}
Hawkes, K., Smith, K.~R., and Robson, S.~L. (2009).
\newblock {Mortality and fertility rates in humans and chimpanzees: How
  within-species variation complicates cross-species comparisons}.
\newblock {\em American Journal of Human Biology}, 21(4):578--586.

\bibitem[Hill et~al., 2001]{144}
Hill, K., Boesch, C., Goodall, J., Pusey, A., Williams, J., and Wrangham, R.
  (2001).
\newblock {{M}ortality rates among wild chimpanzees}.
\newblock {\em Journal of Human Evolution}, 40(5):437--450.

\bibitem[Hill and Hurtado, 1996]{142}
Hill, K. and Hurtado, A. (1996).
\newblock {\em Ache life history: The ecology and demography of a foraging
  people}.
\newblock Aldine de Gruyter, New York.

\bibitem[Howell, 1979]{141}
Howell, N. (1979).
\newblock {\em {Demography of the Dobe !Kung}}.
\newblock Academic Press, New York, 1 edition.

\bibitem[Kachel et~al., 2010]{115}
Kachel, A.~F., Premo, L.~S., and Hublin, J. (2010).
\newblock Grandmothering and natural selection.
\newblock {\em Proceedings of the Royal Society of London B: Biological
  Sciences}.

\bibitem[Kachel et~al., 2011]{113}
Kachel, A.~F., Premo, L.~S., and Hublin, J. (2011).
\newblock Grandmothering and natural selection revisited.
\newblock {\em Proceedings of the Royal Society of London B: Biological
  Sciences}, 278(1714):1939--1941.

\bibitem[Kim et~al., 2012]{117}
Kim, P.~S., Coxworth, J.~E., and Hawkes, K. (2012).
\newblock Increased longevity evolves from grandmothering.
\newblock {\em Proceedings of the Royal Society of London B: Biological
  Sciences}.

\bibitem[Kim et~al., 2014]{116}
Kim, P.~S., McQueen, J.~S., Coxworth, J.~E., and Hawkes, K. (2014).
\newblock Grandmothering drives the evolution of longevity in a probabilistic
  model.
\newblock {\em Journal of Theoretical Biology}, 353:84--94.

\bibitem[Kokko and Jennions, 2014]{131}
Kokko, H. and Jennions, M.~D. (2014).
\newblock {{T}he relationship between sexual selection and sexual conflict}.
\newblock {\em Cold Spring Harbor Perspectives in Biology}, 6(9):a017517.

\bibitem[Lahdenper\"{o} et~al., 2012]{122}
Lahdenper\"{o}, M., Gillespie, D. O.~S., Lummaa, V., and Russell, A.~F. (2012).
\newblock Severe intergenerational reproductive conflict and the evolution of
  menopause.
\newblock {\em Ecology Letters}, 15(11):1283--1290.

\bibitem[Lee, 2008]{128}
Lee, R. (2008).
\newblock Sociality, selection, and survival: Simulated evolution of mortality
  with intergenerational transfers and food sharing.
\newblock {\em Proceedings of the National Academy of Sciences},
  105(20):7124--7128.

\bibitem[Levitis et~al., 2013]{126}
Levitis, D.~A., Burger, O., and Lackey, L.~B. (2013).
\newblock {{T}he human post-fertile lifespan in comparative evolutionary
  context}.
\newblock {\em Evolutionary Anthropology}, 22(2):66--79.

\bibitem[Levitis and Lackey, 2011]{140}
Levitis, D.~A. and Lackey, L.~B. (2011).
\newblock {{A} measure for describing and comparing post-reproductive lifespan
  as a population trait}.
\newblock {\em Methods in Ecology and Evolution}, 2(5):446--453.

\bibitem[Mitteldorf and Wilson, 2000]{165}
Mitteldorf, J. and Wilson, D.~S. (2000).
\newblock {{P}opulation viscosity and the evolution of altruism}.
\newblock {\em Journal of Theoretical Biology}, 204(4):481--496.

\bibitem[O'Connell et~al., 1999]{158}
O'Connell, J.~F., Hawkes, K., and Blurton~Jones, N.~G. (1999).
\newblock Grandmothering and the evolution of homo erectus.
\newblock {\em Journal of Human Evolution}, 36(5):461 -- 485.

\bibitem[Pavard and Branger, 2012]{118}
Pavard, S. and Branger, F. (2012).
\newblock {{E}ffect of maternal and grandmaternal care on population dynamics
  and human life-history evolution: a matrix projection model}.
\newblock {\em Theor Popul Biol}, 82(4):364--376.

\bibitem[Peccei, 2001]{147}
Peccei, J.~S. (2001).
\newblock Menopause: Adaptation or epiphenomenon?
\newblock {\em Evolutionary Anthropology: Issues, News, and Reviews},
  10(2):43--57.

\bibitem[Robbins, 1995]{145}
Robbins, M.~M. (1995).
\newblock A demographic analysis of male life history and social structure of
  mountain gorillas.
\newblock {\em Behaviour}, 132(1/2):21--47.

\bibitem[Sear and Mace, 2008]{167}
Sear, R. and Mace, R. (2008).
\newblock Who keeps children alive? a review of the effects of kin on child
  survival.
\newblock {\em Evolution and Human Behavior}, 29(1):1--18.

\bibitem[Sear et~al., 2000]{121}
Sear, R., Mace, R., and McGregor, I.~A. (2000).
\newblock Maternal grandmothers improve nutritional status and survival of
  children in rural gambia.
\newblock {\em Proceedings of the Royal Society of London B: Biological
  Sciences}, 267(1453):1641--1647.

\bibitem[Sugiyama, 1994]{148}
Sugiyama, Y. (1994).
\newblock Age-specific birth rate and lifetime reproductive success of
  chimpanzees at bossou, guinea.
\newblock {\em American Journal of Primatology}, 32(4):311--318.

\bibitem[Taylor, 1992a]{163}
Taylor, P. (1992a).
\newblock {Altruism in viscous populations -- an inclusive fitness model}.
\newblock {\em Evolutionary Ecology}, 6(4):352--356.

\bibitem[Taylor, 1992b]{164}
Taylor, P. (1992b).
\newblock {Inclusive Fitness in a Homogeneous Environment}.
\newblock {\em Proceedings: Biological Sciences}, 249(1326):299--302.

\bibitem[te~Velde and Pearson, 2002]{123}
te~Velde, E.~R. and Pearson, P.~L. (2002).
\newblock {{T}he variability of female reproductive ageing}.
\newblock {\em Human Reproduction Update}, 8(2):141--154.

\bibitem[Walker et~al., 2006]{149}
Walker, R., Gurven, M., Hill, K., Migliano, A., Chagnon, N., De~Souza, R.,
  Djurovic, G., Hames, R., Hurtado, A.~M., Kaplan, H., Kramer, K., Oliver,
  W.~J., Valeggia, C., and Yamauchi, T. (2006).
\newblock {{G}rowth rates and life histories in twenty-two small-scale
  societies}.
\newblock {\em American Journal of Human Biology}, 18(3):295--311.

\bibitem[West et~al., 2002]{161}
West, S.~A., Pen, I., and Griffin, A.~S. (2002).
\newblock {{C}ooperation and competition between relatives}.
\newblock {\em Science}, 296(5565):72--75.

\bibitem[Williams, 1957]{119}
Williams, G.~C. (1957).
\newblock Pleiotropy, natural selection, and the evolution of senescence.
\newblock {\em Evolution}, 11(4):398--411.

\bibitem[Wilson et~al., 1992]{166}
Wilson, D.~S., Pollock, G.~B., and Dugatkin, L.~A. (1992).
\newblock Can altruism evolve in purely viscous populations?
\newblock {\em Evolutionary Ecology}, 6(4):331--341.

\end{thebibliography}

\end{document}